\def\etal{{\frenchspacing et al. }}
\def\sect{\frenchspacing Section }

\def\fig{\frenchspacing Fig. }
\def\tab{\frenchspacing Table }
   
\documentclass[usenatbib]{mn2e} 
\usepackage{txfonts} 
\usepackage{graphicx}

\title{Low-order multipole maps of cosmic microwave background anisotropy derived from \emph{WMAP}}

\author[P. Bielewicz, K.M. G\'orski
and A.J. Banday]{P. ~Bielewicz,$^{1,2}$ 
\thanks{E-mail: pbielew@fuw.edu.pl}
K.M. ~G\'orski$^{2,3}$ and A.J. ~Banday$^{4}$\\
$^{1}$ Institute of Theoretical Physics, Warsaw
University, ul. Ho\.za 69, PL-00-681 Warsaw, Poland \\ 
$^{2}$ Warsaw University Observatory, Aleje Ujazdowskie 4, 00-478 Warsaw, Poland\\
$^{3}$ JPL, M/S 169/327, 4800 Oak Grove Drive, Pasadena, CA 91109, USA\\
$^{4}$ Max-Planck-Institut fuer Astrophysik, Karl-Schwarzschild-Str.\,1, Postfach
1317, Garching bei Muenchen D-85741, Germany}

\begin{document}
 
\date{Accepted 2004 September 10. Received 2004 July 31; in original form 2004 April 29}

\volume{355}
\pagerange{1283-1302} \pubyear{2004}

\maketitle
   
\begin{abstract}
Recent analyses of the \emph{Wilkinson Microwave Anisotropy Probe}
(\emph{WMAP}) data have suggested that the low-order  
multipoles of the cosmic microwave background (CMB) anisotropy distribution show cosmologically
interesting and unexpected morphologies and amplitudes. 
In this paper, we apply a power equalization (PE) filter to the 
high latitude \emph{WMAP} data in order to reconstruct these low-$\ell$
multipoles free from the largest Galactic foreground modelling
uncertainties in the Galactic plane. 
The characteristic spatial distributions of the modes of order $\ell=2,3,4,5$
have been determined as a function of frequency, sky coverage and 
two methods for foreground correction: using the template-based corrections
of \citet{bennet4} and by a simple linear projection scheme assuming
the spectral dependence of the foreground components.
Although the derived multipole maps are statistically consistent with 
previous estimates from \citet*{tegmark} (TOH) and \citet{efstathiou},
our analyses suggest that the K and Ka frequency bands
remain significantly contaminated by residual foreground emission for the \emph{WMAP}
Kp2 mask. However, the $\ell=3,4,5$ multipole maps for the Q, V and W
channels indicate that, after foreground cleaning, these multipoles are
dominated by the CMB anisotropy component.
We confirm the \citet{tegmark} result that the octopole does 
indeed show structure in which its hot and cold spots are centred on a single 
plane in the sky, and show further that this is very stable with respect to the applied mask
and foreground correction. 
The estimated quadrupole is much less stable showing non-negligible dependence on
the Galactic foreground correction. Including these uncertainties is likely to weaken 
the statistical significance of the claimed alignment between the quadrupole and octopole.
Nevertheless, these anisotropy patterns are also present in the \emph{COBE}-DMR data,
and are unlikely to be associated with instrumental systematic artifacts.

\end{abstract}
\begin{keywords}
cosmic microwave background--cosmology: observations
\end{keywords}

    
\section{Introduction}
In the standard cosmological interpretation, the distribution and 
statistical properties of the cosmic microwave
background (CMB) temperature anisotropies reflect the properties of the Universe
approximately 300 000 years after the big bang. 
The simplest inflationary models of the origin of the primordial
density fluctuations predict statistically isotropic and Gaussian
fluctuations for the CMB.
Thus, in the conventional approach, where the temperature anisotropy field, $\Delta
T(\theta,\phi)$, is expanded as a sum over the spherical harmonic 
basis on the sphere
\begin{equation} \label{dtemp}
\Delta
T(\theta,\phi)=\sum_{\ell=1}^\infty \sum_{m=-\ell}^\ell a_{\ell m} Y_{\ell
m}(\theta,\phi) \
\end{equation} 
the decomposition coefficients, $a_{\ell m}$, are
independent, Gaussian distributed stochastic variables. Then, as a result of statistical isotropy, their  
variances $C_\ell=\left<|a_{\ell m}|^2 \right>$ depend only on $\ell$.
The validity of these predictions can be constrained by direct observation of the
CMB anisotropy.

The sum over $m$ for a given $\ell$, $\Delta T_\ell (\theta, \phi)$,
is then a multipole map of order $\ell$, and typically represents the
CMB fluctuations on angular scale $\sim 1/\ell$. 
The lowest multipoles of the CMB anisotropy, in particular the dipole ($\ell=1$), quadrupole
($\ell=2$) and octopole ($\ell=3$), are especially important for
studying the homogeneity and isotropy of the Universe after the last scattering of the CMB photons 
\citep*{maartens}. The measurement of these low-order 
multipoles requires observations over a large fraction of the sky,
as provided initially by the \emph{COBE}-DMR experiment, and  
more recently by the \emph{Wilkinson Microwave Anisotropy Probe} (\emph{WMAP}) satellite.
The dipole is generally interpreted as being dominated by a kinetic effect related to the peculiar
motion of the Earth with respect to the CMB rest frame, so cosmologically relevant
studies usually consider multipoles only from the quadrupole upwards.
The first measurement of the quadrupole by \emph{COBE}-DMR suggested an intriguingly low value for 
its amplitude, and led to vigorous discussion about the credibility of its estimation  
(\citealt{smoot}; \citealt{bennet2}; \citealt{stark}; \citealt{gould}; \citealt{kogut}). 
In particular, doubts centred on limitations imposed by systematic effects, 
instrumental noise, and non-cosmological foreground contributions due
to local astrophysical emission. 
Nevertheless, the low amplitude of the quadrupole was
recently confirmed by high precision measurements from the \emph{WMAP} satellite \citep{bennet3}
after the application of a refined analysis of Galactic foreground
contamination. 
 
Subsequently, \citet{tegmark} (hereafter TOH) have pointed out that
the quadrupole and octopole appear rather planar, with most of the hot
and cold spots centred on a single plane in the sky. Moreover, the two
planes appear rather aligned.  Other studies have also demonstrated
the presence of intriguing correlations (\citealt*{copi};
\citealt{schwarz}) and anomalies (\citealt{eriksen2};
\citealt{larson}; \citealt{vielva}) of the low-order \emph{WMAP}
multipoles.  Some of these features could be explained by a specific
non-standard topology for the Universe, although a broad class of
finite universe models, including the Poincar\'{e} Dodecahedron model
proposed by \citet{luminet}, was recently ruled out by
\citet{cornish}.  However, many of these previous analyses have
determined the multipole amplitudes on the full sky anisotropy as
measured by either the internal linear combination
\citep[ILC, ][]{bennet4}\footnote{Indeed, it should be noted that such
analyses explicitly ignore the admonition of the \emph{WMAP} team that
the map is not suitable for cosmological analysis.} or TOH
map. Importantly, \citet{eriksen} have shown that residual Galactic foreground
uncertainties for the ILC map in the Galactic plane can be at the
level of approximately 50 $\mu \rm{K}$,
whilst the analysis of \citet{slosar} strongly suggests that much of
the detected alignment is indeed related to that part of the data most
contaminated by (residual) Galactic foreground. 

The goal of this paper is to estimate the low-order \emph{WMAP}
multipoles using a technique which allows the explicit determination
of their reconstruction uncertainties, and subsequently to clarify
whether the unexpected features of the quadrupole and octopole are
significant. Some of these uncertainties are related to Galactic
foreground emission.  In what follows, we consider that the estimation
of CMB anisotropy based on the \emph{WMAP} maps is not credible when
including data from the Galactic plane, and we will omit this region
using a set of increasingly aggressive masks.  Even after this data
excision, the remaining high-latitude Galactic foreground emission may
compromise our multipole reconstructions.  Therefore, we adopt two
independent methods of foreground removal.  In the first, we use three
templates for the Galactic synchrotron, free-free and dust (including the
so-called anomalous dust-correlated emission), scaled to the \emph{WMAP} frequencies by
the weights provided by the\emph{WMAP} team and subsequently
subtracted from the individual observation maps. This treatment is
completely consistent with that of the \emph{WMAP} team and has
largely been adopted as the standard correction by the CMB community at-large. 
For the second method, we do not use templates, but simply take linear
combinations of the \emph{WMAP} data in five frequency bands designed
to project out the CMB component after assuming spectral index values
for the individual foreground components.  More details about the
methods and their limitations will be found in the appropriate
sections of the paper.  The multipoles determined from the
foreground-corrected data will also be compared to those computed from
the ILC and TOH cleaned maps, to which we also apply masking of the
Galactic plane. Because these were derived using a different technique
for removing the Galaxy, this provides yet further constraints on the
foreground contributions.  Note that it is not our intention to
provide a definitive model of Galactic emission, but simply to
investigate the stability of the inferred low-$\ell$ anisotropy
patterns to different assumptions.  If the multipoles are stable to
foreground correction then we will infer that the anisotropy is likely
to be cosmological in origin.

\section{The problem} \label{problem}  
\label{sec:filters}
One of the major goals of measurements of the CMB radiation
is to allow for the construction of a map of the
anisotropy. To facilitate this, the measurements are discretized into $N_{\rm{pix}}$ pixels, which
can be written as a $N_{\rm{pix}}$-dimensional data vector $\mathbf{d}$. 
This is related to the true sky signal $\mathbf{x}$ (itself the sum of
the true CMB anisotropy field $\mathbf{x_{CMB}}$ and the foreground
contribution $\mathbf{x_{fg}}$) by
\begin{equation} \label{yxn1}
\mathbf{y}=\mathbf{B}\mathbf{x}+\mathbf{n} \ ,
\end{equation} 
where the matrix $\mathbf{B}$ encodes the effect of beam and pixel smoothing, while the additional 
term $\mathbf{n}$ represents the contribution from detector noise. 
The foreground contamination can initially be minimized by masking out those regions of the sky
with strong foreground emission, e.g. as a result of the Galactic plane or point sources. The remainder of the map can then
be corrected for the high-latitude Galactic (and extragalactic if necessary) emission. 
The vector $\mathbf{d}$ is then considered to denote a cut sky, foreground corrected map.  

In the spherical harmonic, $Y_{\ell m}$, space,
the vector\footnote{Mapping from an index pair 
$(\ell,m)$ into a single index $i$ is given by \mbox{$i=\ell^2+\ell +m+1$}} of 
decomposition coefficients of the map on the cut sky, $\tilde{\mathbf{a}}$, is related to the corresponding
vector on the full sky, $\mathbf{a}$, by 
\begin{equation} \label{yxn2}
\tilde{\mathbf{a}}=\mathbf{K}\cdot \mathbf{B} \cdot \mathbf{a}+\mathbf{n}_c \ ,
\end{equation}
where $\mathbf{n}_c$ denotes the noise vector on the cut sky 
and $\mathbf{B}$ now consists of the appropriate harmonic coefficients of the beam window 
function. We will assume that the beam profile is azimuthally
symmetric. Then $\mathbf{B}$ is a diagonal matrix dependent only
on $\ell$ and consisting of the weights $w_\ell^{\rm{beam}}$. To simplify notation, hereafter
$\mathbf{a}$ will denote the vector multiplied by $\mathbf{B}$.
$\mathbf{K}$ is referred to as the coupling matrix and its $(i,j)$-th component is given by
\begin{equation}
K_{i(\ell,m),j(\ell',m')}=\int_{\rm{cut\ sky}} Y_{\ell m}^\ast (\hat{\mathbf{n}}) 
Y_{\ell' m'}^{}(\hat{\mathbf{n}}) d \Omega_{\hat{\mathbf{n}}} \ .
\end{equation} 

Our aim is to determine the full sky representation of the 
anisotropy written in terms of the coefficients $\mathbf{a}$ using the above relation.
Unfortunately, we do not know the noise contribution to this specific
set of observations, only a statistical 
description of the ensemble properties, and moreover 
the sky cut causes the coupling matrix $\mathbf{K}$ to be singular. 
This means that there is no information 
on those anisotropies that project onto modes in the spherical harmonic basis where
the only structure lies within the sky cut. Hence it is impossible to reconstruct all of the modes from 
the $\tilde{\mathbf{a}}$. However, for low-order multipoles, small sky cuts and a high signal-to-noise
ratio, a good approximation is to simply truncate the vectors and coupling matrix at some value of the index $i(\ell,m)$  
then reconstruct the multipole coefficients by inverting the non-singular 
truncated matrix $\mathbf{K}$. This method was applied 
to the first year \emph{WMAP} data in \citet{efstathiou2}, and will hereafter be
referred to as direct inversion.

Direct inversion is a special case of a more general technique for estimation 
of the true sky signal by a linear transformation of the data vector 
$\tilde{\mathbf{a}}$ 
\begin{equation}  
\hat{\mathbf{a}}=\mathbf{F}\cdot \tilde{\mathbf{a}} \ ,
\end{equation}     
where matrix $\mathbf{F}$ acts as a filter and satisfies certain conditions imposed on the solution 
$\hat{\mathbf{a}}$. In this paper, we will consider 
both \citet{wiener} and power equalization (PE) filters, and utilise the
detailed formalism developed by \citet{gorskimoriond}
for the application of these filters to CMB observations.

\subsection{Wiener filter}
The Wiener filter \citep{wiener} is constructed from the requirement that the mean square deviation
between the filtered, $\hat{\mathbf{a}}$, and the true signal, $\mathbf{a}$,
\mbox{$\epsilon=\left< (\hat{\mathbf{a}}-\mathbf{a})^T \cdot
(\hat{\mathbf{a}}-\mathbf{a})\right>=\textrm{tr} \left<(\hat{\mathbf{a}}-\mathbf{a}) \cdot
(\hat{\mathbf{a}}-\mathbf{a})^T \right>$}, is minimized. Then we obtain
\begin{equation} \label{a}
\mathbf{F}_W=\mathbf{S} \cdot \mathbf{K}^T
\cdot \left( \mathbf{K}\cdot\mathbf{S} \cdot \mathbf{K}^T+
\mathbf{N} \right)^{-1}\ ,
\end{equation}
(see, e.g., \citealt{bunn}; \citealt{zaroubi}) where $\mathbf{S}\equiv
\left<\mathbf{a} \cdot \mathbf{a}^T \right>$ and $\mathbf{N}\equiv \left<\mathbf{n}_c
\cdot \mathbf{n}^T_c \right>$ are covariance matrices of the signal
and noise, respectively. One should notice that this filter 
practically annihilates those components of the data which are assessed to be 
noise dominated. Non-cosmological modes can be removed from the 
analysis by assumption that their power spectrum in the matrix
$\left( \mathbf{K}\cdot\mathbf{S} \cdot \mathbf{K}^T+
\mathbf{N} \right)^{-1}$ is very large. In practice, the monopole and dipole 
are such terms, because for differencing experiments such as \emph{WMAP} the
monopole may correspond to an unphysical quantity related to the map-making process, 
whereas the dipole is attributed to the motion of the Earth in the Universe.
For negligible noise the filter is essentially equivalent to the direct inversion method and does 
not depend on the assumed signal covariance matrix. However, we would like to explicitly take into account 
the correlations induced by the cut between the lower order modes we are interested in reconstructing
with the higher order modes present in the data. These correlations are non-negligible and play the role 
of noise in the Wiener filter.

For further consideration, it will be useful to introduce a new orthonormal
basis of functions on the cut sky. 
A method of construction such of such orthonormal functions,
$\bpsi$, was proposed by \citet{gorski}.

In this new basis, the vector of decomposition coefficients of the cut sky map, $\mathbf{c}$, is 
related to vector $\mathbf{a}$ by
\begin{equation} \label{ca_relation} 
\mathbf{c}=\mathbf{L}^T \cdot \mathbf{a}+\mathbf{n}_\psi \ ,
\end{equation}
where $\mathbf{L}$ is the matrix derived by the 
Choleski decomposition of the coupling matrix and $\mathbf{n}_\psi$
is the vector of noise coefficients in the $\bpsi$ basis. 
Because $\mathbf{L}$ is a lower triangular matrix, $\mathbf{c}$ can be
decomposed into a part dependent on both lower $\mathbf{a}_\mathcal{L}$ and higher
$\mathbf{a}_\mathcal{H}$ order modes 
\begin{equation} \label{b} 
\mathbf{c}_\mathcal{L}=\mathbf{L}_{\mathcal{LL}}^T\cdot
\mathbf{a}_\mathcal{L}+\mathbf{L}_{\mathcal{HL}}^T\cdot 
\mathbf{a}_\mathcal{H}+\mathbf{n}_{\psi, \mathcal{L}} \ ,
\end{equation}
and a part dependent only on higher order modes
\begin{equation} \label{c} 
\mathbf{c}_\mathcal{H}=\mathbf{L}_{\mathcal{HH}}^T\cdot 
\mathbf{a}_\mathcal{H}+\mathbf{n}_{\psi, \mathcal{H}} \ .
\end{equation}
The subscripts $\mathcal{L}$ and $\mathcal{H}$ denote 
the range of indices \mbox{$i=1,\dots,(\ell_{\rm{rec}}+1)^2$} and
\mbox{$i=(\ell_{\rm{rec}}+1)^2+1,\dots, (\ell_{\rm{max}}+1)^2$}
corresponding to those modes which are to be reconstructed with the filter (multipoles
in the range from 1 to $\ell_{\rm{rec}}$) and the remainder
of the modes used in the analysis (multipoles in the range from
$\ell_{\rm{rec}}+1$ to $\ell_{\rm{max}}$), respectively. 
It should be noticed that only the first part of the 
vector $\mathbf{c}$ contains information about lower order modes 
$\mathbf{a}_\mathcal{L}$, so to estimate them we will use only this part of
the vector. 
This is a significant difference as compared to the vector
$\tilde{\mathbf{a}}$ 
where information about the lower order modes is also contained in the
second part, $\tilde{\mathbf{a}}_\mathcal{H}$.

Taking $\mathbf{c}_\mathcal{L}$ as a data vector, estimation of the lower order
modes will be given by $\hat{\mathbf{a}}_\mathcal{L}=\mathbf{F} \cdot
\mathbf{c}_\mathcal{L}$. For the Wiener filter we have
\begin{equation} \label{wiener_fil} 
\mathbf{F}_W=\mathbf{S}_\mathcal{LL} \cdot \mathbf{L}_{\mathcal{LL}} \cdot
\mathbf{C}^{-1} \ , 
\end{equation}  
where $\mathbf{C}=\mathbf{L}_{\mathcal{LL}}^T \cdot \mathbf{S}_\mathcal{LL}
\cdot \mathbf{L}_{\mathcal{LL}}+ \mathbf{L}_{\mathcal{HL}}^T \cdot
\mathbf{S}_\mathcal{HH} \cdot \mathbf{L}_{\mathcal{HL}}+\mathbf{N}_\psi$ and 
$\mathbf{S}_\mathcal{LL}$, $\mathbf{S}_\mathcal{HH}$ are the appropriate covariance
matrices of the CMB modes corresponding to the lower and higher
multipoles, respectively. $\mathbf{N}_\psi$ is the noise covariance matrix
in the $\bpsi$-basis.

\subsection{Power equalization filter}
The PE filter is defined by the requirement that $\mathbf{F}$ is chosen such that
\begin{equation} \label{pe_cond} 
\left<
\hat{\mathbf{a}} \cdot \hat{\mathbf{a}}^T \right> = \mathbf{S}_\mathcal{LL},
\ \ \textrm{i.e.}\ \ \mathbf{F}_{PE}\cdot \mathbf{C} \cdot \mathbf{F}_{PE}^T= \mathbf{S}_\mathcal{LL}
\ .
\end{equation}
Using the Choleski decomposition of the relevant matrices, 
we can construct the PE filter as follows:
\begin{eqnarray} 
\mathbf{S}_\mathcal{LL}^{-1}=\mathbf{L}_{\mathbf{S}_\mathcal{LL}^{-1}}^{}\cdot
\mathbf{L}_{\mathbf{S}_\mathcal{LL}^{-1}}^T, \ \ & 
\mathbf{\Gamma}_{\mathbf{S}_\mathcal{LL}^{-1}}^{}=\mathbf{L}_{\mathbf{S}_\mathcal{LL}^{-1}}^{-1},
\ \ & 
\mathbf{S}_\mathcal{LL}=\mathbf{\Gamma}_{\mathbf{S}_\mathcal{LL}^{-1}}^T
\cdot \mathbf{\Gamma}_{\mathbf{S}_\mathcal{LL}^{-1}}^{}, 
\end{eqnarray}

\begin{eqnarray}
 \mathbf{C}^{-1} =\mathbf{L}_{\mathbf{C}^{-1}}^{} \cdot 
\mathbf{L}_{\mathbf{C}^{-1}}^T,\  \ 
  & \mathbf{\Gamma}_{\mathbf{C}^{-1}}^{}  =
\mathbf{L}_{\mathbf{C}^{-1}}^{-1},  \ \   
  & \mathbf{C} = \mathbf{\Gamma}_{\mathbf{C}^{-1}}^T \cdot 
\mathbf{\Gamma}_{\mathbf{C}^{-1}}^{},  
\end{eqnarray}
and as a result we obtain an upper triangular matrix:
\begin{equation}
\mathbf{F}_{PE}=\mathbf{\Gamma}_{\mathbf{S}_\mathcal{LL}^{-1}}^T \cdot 
\mathbf{L}_{\mathbf{C}^{-1}}^T \ ,
\end{equation} 
where the matrices $\mathbf{S}_\mathcal{LL}$ and $\mathbf{C}$ are the same as in
the Wiener filter case. 

The condition (\ref{pe_cond}) means that on average over many applications the PE
filter renders filtered data whose statistically most likely power spectrum matches that of the underlying
signal. Because the PE filter is an upper triangular matrix, the non-cosmological
monopole and dipole terms can be explicitly eliminated from the analysis.
Then, the restored modes (with $i > 4$) are not contaminated by the monopole and dipole contributions.
\\

\subsection{Errors}
The errors of estimation are determined on the basis of the diagonal terms
of the covariance matrix, defined as 
\mbox{$\mathbf{M}=\left< (\mathbf{a}_\mathcal{L}-\hat{\mathbf{a}}_\mathcal{L}) 
(\mathbf{a}_\mathcal{L}-\hat{\mathbf{a}}_\mathcal{L})^T
\right>$}. Thus, we have 
\begin{eqnarray} \label{cov_matrix} 
\mathbf{M}& =& \left(\mathbf{F}\cdot
\mathbf{L}_\mathcal{LL}^T-\mathbf{I}\right) \cdot \mathbf{S}_\mathcal{LL}
\cdot \left(\mathbf{F}\cdot
\mathbf{L}_\mathcal{LL}^T-\mathbf{I}\right)^T + \nonumber \\
& & +\mathbf{F}\cdot (\mathbf{L}_{\mathcal{HL}}^T \cdot
\mathbf{S}_\mathcal{HH} \cdot \mathbf{L}_{\mathcal{HL}}) \cdot \mathbf{F}^T
+\mathbf{F}\cdot \mathbf{N}_\psi \cdot \mathbf{F}^T \ ,
\end{eqnarray}
where $\mathbf{I}$ is the identity matrix. This is a sum of three components:
the first is a result of correlations between the recovered low-order multipoles, 
the second is induced by correlations with higher order multipoles and the last 
is generated by noise. 

\subsection{General considerations}
In order to construct the filters we have to assume the form of both 
the signal and noise covariance matrices.
We will make the usual assumption that both the CMB and noise components are Gaussian
stochastic variables. The
variances for individual modes of the CMB anisotropy, $C_\ell=\left<a^2_{\ell m} \right>$, are
uniquely expressed as integrals over the power spectrum of the matter density fluctuations
and depend only on $\ell$ as a result of the statistical isotropy of the CMB temperature
field. The rms noise level in pixel $p$,
\mbox{$\sigma(p)_{\rm{noise}}=\sigma_o/\sqrt{N_{\rm{obs}}(p)}$}, depends on quantities specific
to a given instrument and scan strategy: the number of observations in pixel,
$N_{\rm{obs}}(p)$, and the rms noise per observation, $\sigma_o$. 
Dependence on the assumed power spectrum
might seem to be a disadvantage of these filtering methods. However, we will see that in
the case of the \emph{WMAP} data, the multipole estimation does 
not depend significantly on the assumed power spectrum. 

\section{Data} \label{data} 
The \emph{WMAP} satellite observes the sky in five frequency bands denoted \emph{K}, \emph{Ka}, 
\emph{Q}, \emph{V} and \emph{W}, centred on the frequencies of 22.8, 33.0, 40.7, 60.8 and 93.5
GHz, respectively. The maps\footnote{Available at http://lambda.gsfc.nasa.gov} 
are pixelized in the HEALPix\footnote{http://www.eso.org/science/healpix} 
scheme \citep*{healpix} with a resolution parameter
$N_{\rm{side}}=512$, corresponding to 3 145 728 pixels with a 
pixel size of $\sim 7$ arcmin. The maps 
are corrected for the Galactic foreground 
using three templates$^{4}$ 
[a radio survey at 408 MHz \citep{haslam}, a
H$\alpha$ map \citep{finkbeiner}, and 
a dust map based on the combined \emph{COBE}-DIRBE and IRAS data
\citep*{fds}] to trace the synchrotron, free-free and dust emission.
\citet{bennet4} have concluded that the combination of these
templates, each scaled to the frequencies in question by a single
factor derived for the high-latitude sky (i.e. making no allowances for
spectral variations in the foreground behaviour), is 
sufficient to account for almost all of the foreground
emission including the anomalous dust-correlated component (irrespective of its
physical origin). Although the \emph{WMAP} team
did undertake an analysis of the data based on the maximum entropy method (MEM) which allowed for spectral index
variations of the foreground components on a pixel-by-pixel basis, the
resultant CMB map was deemed inappropriate for use as a result of complex
noise properties. For all cosmological data analysis, the \emph{WMAP}
team have corrected the data using the template method.

We adopt the coefficients from \citet{bennet4}, which were derived 
by fitting the templates to the data for the Kp2 sky coverage,
and then subtract the appropriately scaled data from the \emph{WMAP}
sky maps\footnote{
In fact, the LAMBDA web provides a set of foreground corrected
maps generated with this recipe. However, a problem was discovered --
see \emph{http://lambda.gsfc.nasa.gov/product/map/IMaps\_cleaned.cfm} --
in that the templates used in this process were convolved with a 
too low $l_{\textrm{max}}$ (the maximum multipole component), 
resulting in ringing around strong point sources. Although we do not
expect this to compromise our results on large angular scales, we
have, nevertheless, generated our own template corrections.}.
For the \emph{Q}, \emph{V} and \emph{W} bands five coefficients were determined: 
an amplitude for  each of the synchrotron and free-free 
components at 41 GHz (and scaled to the other frequencies by 
assuming spectral indices of $-2.7$ and $-2.15$ respectively),
and three for the dust, one at each frequency. The anomalous
dust-correlated emission was thus effectively absorbed into the dust 
template scaling (which also accounts for the thermal contribution).
For the K and Ka channels, the coefficients from table 3 of 
\citet{bennet4} were used.  
The \emph{Q}-, \emph{V}- and \emph{W}-band maps are then combined using inverse-noise-variance weights
to create the corrected co-added map.

We utilize several masks in the work which follows:
the Kp2 mask (eliminating $\sim 15$ per cent of the sky),
the Kp0 mask (eliminating $\sim 23$ per cent of the sky), 
an extended Galactic mask and mask 30. 
The extended Galactic mask (hereafter denoted 20+) is the same
as that used for the \emph{COBE}-DMR data analysis
and is defined as the region with latitude $|b|>20^\circ$ plus custom cut-outs 
in the vicinity of Orion and Ophiuchus \citep{bennet}. 
The broadest mask 30 is defined as the
region with latitude $|b|>30^\circ$. The latter masks were
additionally modified by excluding the point source mask included
in the Kp2 and Kp0 masks, so that they finally eliminate $\sim 38$ and
$\sim 50$ per cent of the sky, respectively.
It should be noted here that changing the Galactic cut applied to the
data could, in principle, require modulation of the scaling coefficients used for the
templates, particularly if the foregrounds do demonstrate strong
variations in spectral index with Galactic latitude. 
However, we proceed with the simple foreground model specified by the 
\emph{WMAP} team and used by them (and elsewhere in the literature)
irrespective of the applied mask.

\section{Choice of $\ell_{\rm{max}}$ and $\ell_{\rm{rec}}$} \label{lmaxlrec}
The choice of $\ell_{\rm{max}}$ and $\ell_{\rm{rec}}$ is based on Monte Carlo (MC) simulations 
of the \emph{WMAP} data, consisting of 
both realizations of the CMB anisotropy and noise. The CMB anisotropy maps were generated as random 
realizations of a Gaussian field with an angular power spectrum 
corresponding to the best-fitting $\Lambda$CDM cosmological model with running spectral 
index\footnote{Available at http://lambda.gsfc.nasa.gov} 
\citep{bennet3,spergel,hinshaw}, convolved with the channel specific
beam window function $w_\ell^{\rm{beam}}$. 
The noise maps were produced from Gaussian random numbers with pixel-dependent
variance corresponding to the properties of the 
corrected co-added map. The simulated maps were studied after masking by the Kp2 mask.

The upper limit for possible values of $\ell_{\rm{max}}$ and $\ell_{\rm{rec}}$ is imposed by
the feasibility of the Choleski decomposition of the coupling matrix. For a given cut, the 
coupling matrix becomes numerically singular for multipoles larger than 
$\ell_{\rm{Choleski}}$ at which point the Choleski decomposition fails.
For the Kp2 mask this multipole is found at $\ell_{\rm{Choleski}} \sim
70$. The studies in this paper are focused mostly on estimation of the 
lowest multipoles ($2 \leq \ell \leq 10$), and particularly on those modes 
which have been shown to have potentially anomalous properties ($2 \leq \ell \leq 5$).
Thus it is sufficient to include only multipoles up to
$\ell_{\rm{max}}=30$. Increasing $\ell_{\rm{max}}$ to 35 or 40 did not appreciably change
the estimated values of these multipoles. This is a direct consequence of the decreasing amplitude of
higher multipoles, scaling roughly as $\ell^{-1}$ ($C_\ell \propto \ell^{-2}$).

\begin{figure*}

\includegraphics[angle=0.,width=0.7\textwidth]{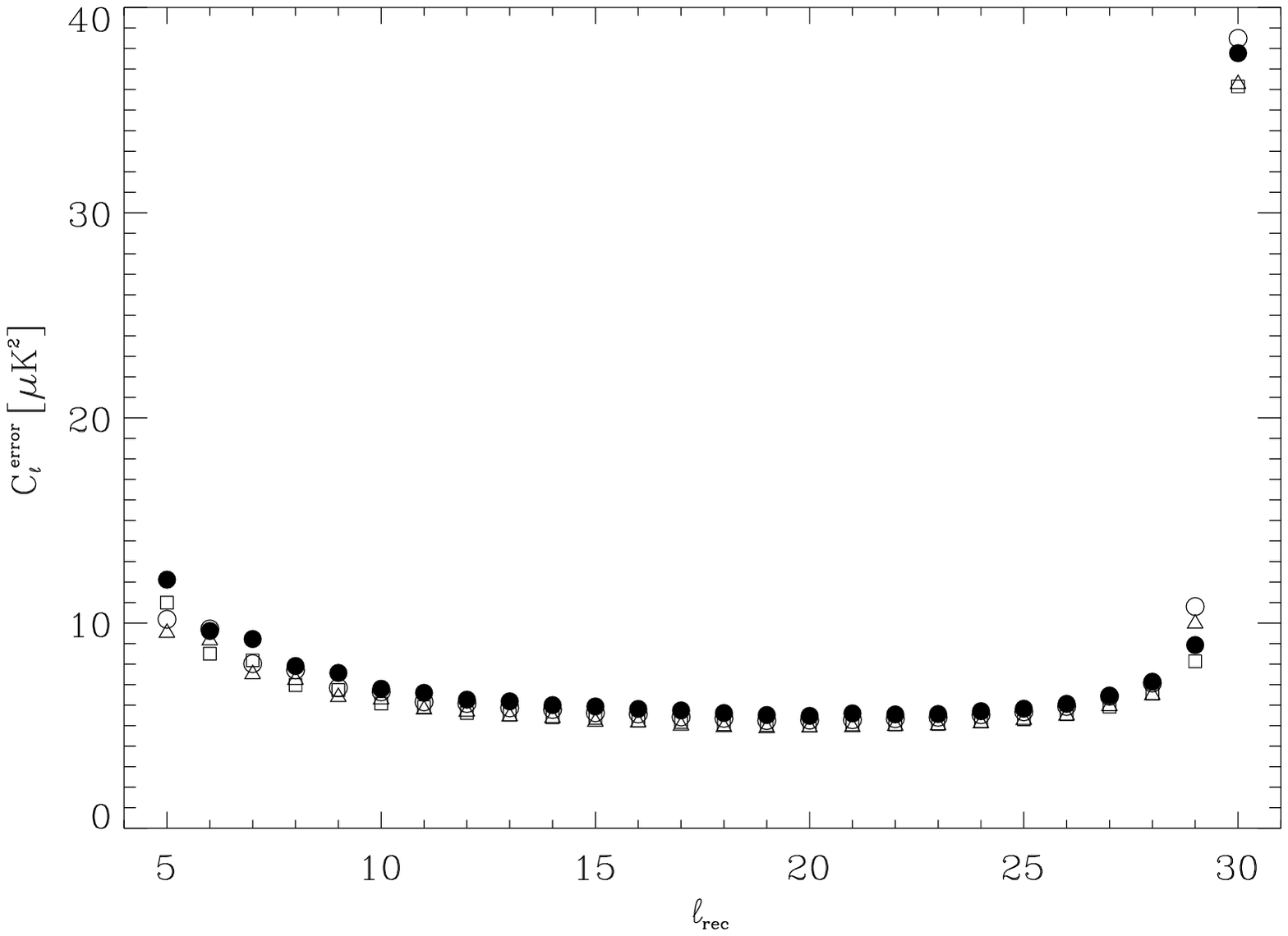}
\caption{The rms of the difference between the input and output
multipole maps as a function of $\ell_{\rm{rec}}$. Filled circle, empty circle, square and triangle
denote $C_\ell^{\rm{error}}$ for multipoles $\ell=2,3,4,5$, respectively.}
\label{clerr}
\end{figure*}

In order to establish a suitable value for $\ell_{\rm{rec}}$,  
1000 MC simulations were generated and the PE filter computed with $\ell_{\rm{max}}=30$ applied to
the simulated maps masked by the Kp2 mask. The power spectrum of the 
difference between the true and restored multipole coefficients over the range $2 \leq \ell \leq 5$
$$C^{\rm{error}}_\ell=\frac{\sum^\ell_{m=-\ell} \left<|a_{\ell m}^{\rm{output}}-a_{\ell
m}^{\rm{input}}|^2\right>}{2\ell+1}\ ,$$ is shown in \fig(\ref{clerr}) as a function of $\ell_{\rm{rec}}$. Notice that for
\mbox{$\ell_{\rm{rec}}=\ell_{\rm{max}}=30$}, the errors are the largest. In this case,
the PE filter does not adequately take into account the aliasing of higher order modes into
the range $\ell \leq \ell_{\rm{rec}}$. Significant errors are seen also for lower values of   
$\ell_{\rm{rec}}$, caused by correlations of the multipoles
$\ell=2,\dots,5$ with modes of order $\ell_{\rm{rec}}+1,
\ell_{\rm{rec}}+2$ which for low values of $\ell_{\rm{rec}}$ have significant amplitudes. 
In what follows we adopt $\ell_{\rm{rec}}=10$; although the flat power
spectrum of the uncertainties over the range $10 \lesssim \ell_{\rm{rec}} \lesssim 25$
allows a larger value for $\ell_{\rm{rec}}$ to be selected, 
later in the paper we will consider the broader
Galactic cut 30 for which the PE filter fails when too large a value for
$\ell_{\rm{rec}}$  is used.

It should be noted that other methods for the determination of the low multipoles
also require the choice of some parameters to facilitate the reconstruction method.
In particular, the direct inversion method of \citet{efstathiou} adopts
two parameters without justification: the order of the multipole at which the coupling matrix
is truncated and the FWHM of the beam for the initial smoothing of the data.

\section{Comparison of the Wiener and PE filters}
\label{comparision_pe_wiener}
\begin{figure*} 
\begin{center}
\includegraphics[angle=0.,width=0.7\textwidth]{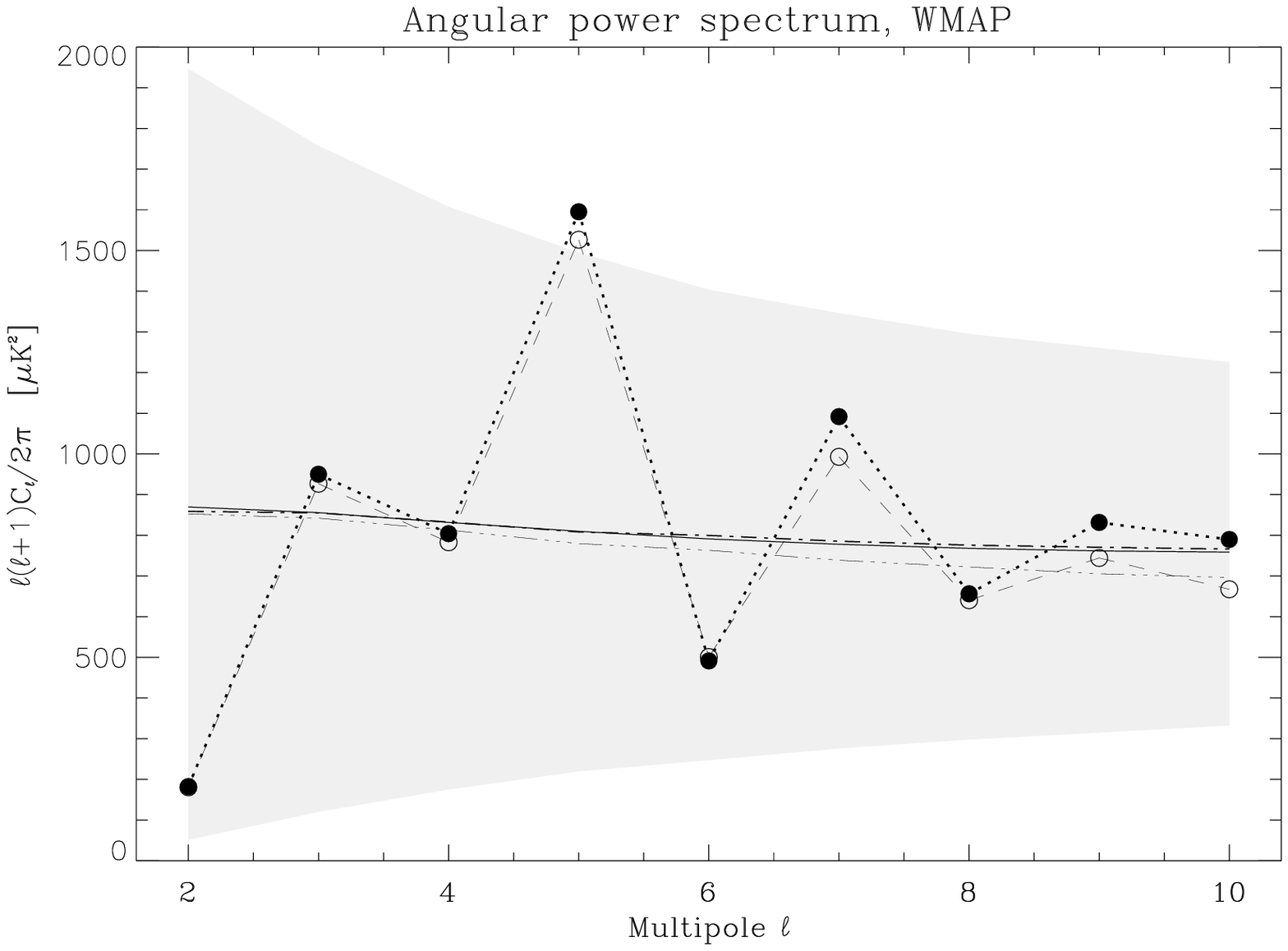}
\caption{Comparison of the power spectra of the \emph{WMAP} map, masked by the Kp2
mask, restored by the Wiener (open circles) and PE (filled circles)
filters. Dash dot dot and dash dot lines correspond to the
averaged power spectra derived from the Wiener and PE filter reconstruction of $10^4$
simulated \emph{WMAP} maps respectively. The smooth curve shows the \emph{WMAP} team
best-fitting $\Lambda$CDM model ($\Omega_\Lambda=0.73$ and $\Omega_m=0.27$)
with running spectral index. The grey band indicates the cosmic variance
errors (95 per cent confidence level).} \label{cl_wmap_wiener_pe}
\end{center} 
\end{figure*}

\fig(\ref{cl_wmap_wiener_pe}) provides a comparison of the \emph{WMAP} power spectra obtained by
the Wiener and PE filtering. In addition, we show the averages over 10 000
\emph{WMAP} map simulations for the power spectra of the restored maps. In both cases the Kp2
mask and filters with $\ell_{\rm{rec}}=10$, $\ell_{\rm{max}}=30$ were applied.

The Wiener and PE filters differ mainly in the recovered power of the multipoles:
there is a systematic suppression of power with increasing $\ell$ for maps 
reconstructed by the Wiener filtering. Conversely, the definition of the PE filter ensures an 
unbiased estimate for the multipole amplitudes. However, the PE filter does not give an optimal
estimation in the minimum variance sense, i.e. an estimate which has the least variance between the
true and restored data, which is the case for the Wiener filter case. 
Nevertheless, we decided to use the PE filter in the remainder of the paper
because for the lowest multipoles and Kp2 mask the results do not differ 
significantly from those derived by the optimal Wiener filter, 
and the unbiased nature of the estimator is considered to be more important.
This is particularly relevant with increasingly aggressive sky cuts where the
Wiener filter demonstrates even more suppression of power.

\section{Comparison of the direct inversion and the PE filtering methods}
\label{comparision}
\begin{figure*} 
\begin{center}
\includegraphics[angle=0.,width=0.7\textwidth]{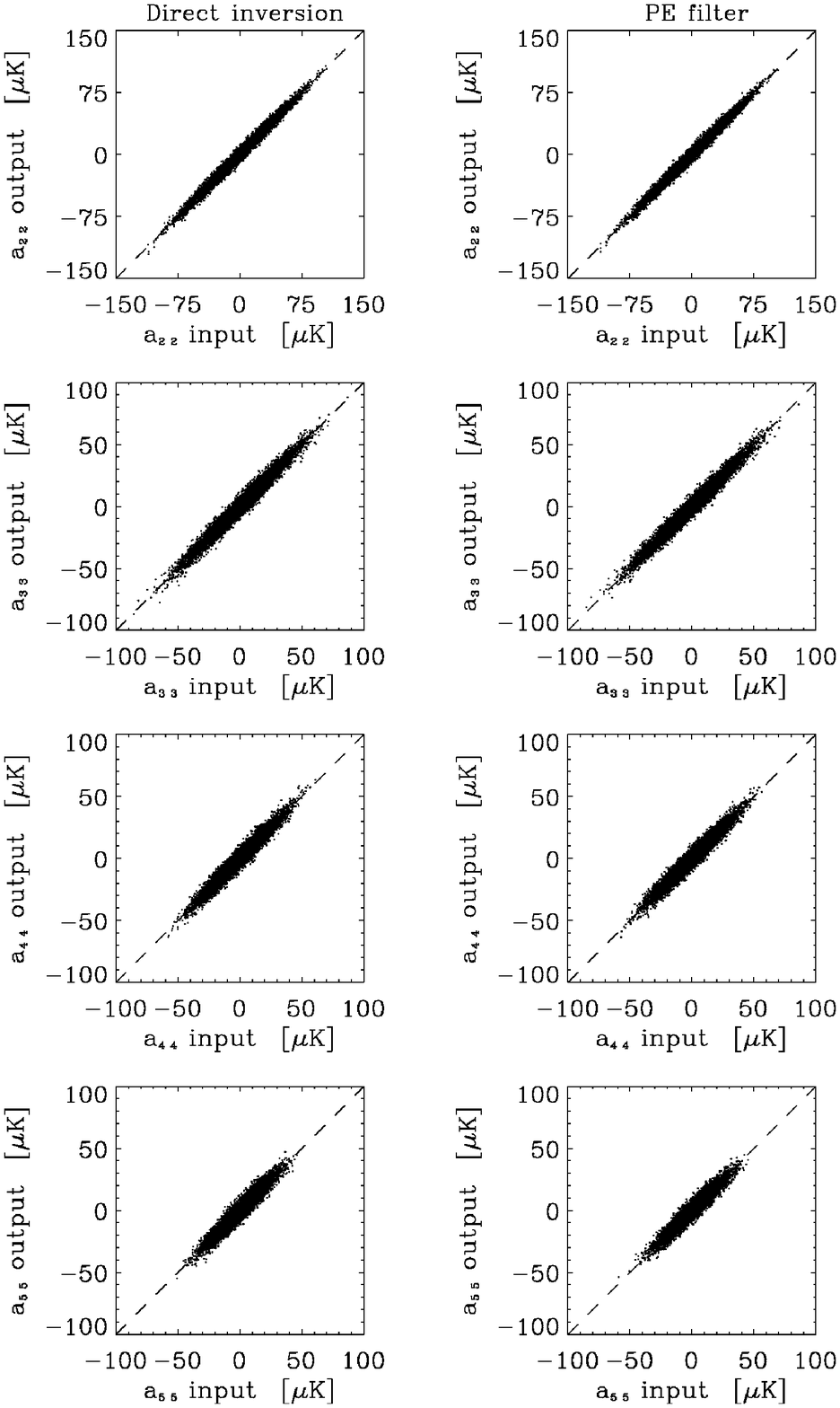}
\caption{Comparison of the accuracy of estimation of the $a_{\ell,\ell}$
modes using the direct inversion and PE filtering reconstruction methods,
determined from simulations with the Kp2 mask. The abscissae give the input values 
of harmonic coefficients used to generate the simulated skies. The ordinates give
the output values from the direct inversion (left column) and PE filtering (right column)
methods. The coupling matrix was truncated at $\ell_{\rm{rec}}=10$ in the direct 
inversion method.} \label{alm_corr}
\end{center} 
\end{figure*}

\begin{figure*}
\begin{center}
\includegraphics[angle=0.,width=0.7\textwidth]{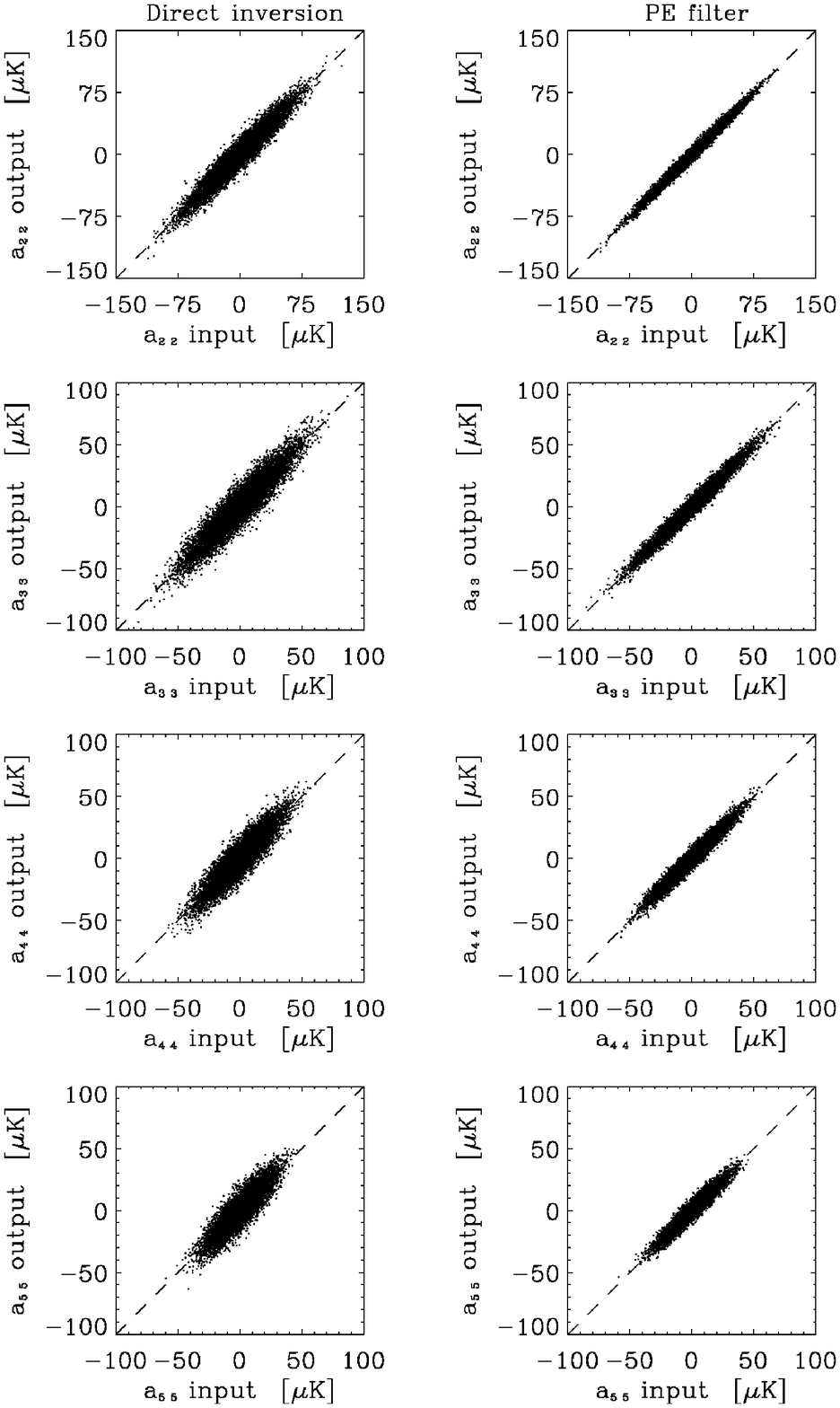}
\caption{The same as Fig. (\ref{alm_corr}) but with the 
coupling matrix truncated at $\ell_{\rm{rec}}=30$ in the direct inversion method.} \label{alm_corr2}
\end{center}
\end{figure*}

10 000 simulations of the \emph{WMAP} data were again used to compare the accuracy of
reconstruction of the low-order multipoles by the direct inversion and the PE
filtering methods. The reconstructions were performed for the Kp2 mask, and
for a PE filter constructed with $\ell_{\rm{rec}}=10$ and 
$\ell_{\rm{max}}=30$. For the direct inversion method the
coupling matrix was evaluated after truncation at either $\ell_{\rm{rec}}=10$ 
or 30. In both cases, following \citet{efstathiou} the input map was smoothed by
a Gaussian beam with FWHM=$7^\circ$. Smoothing improves the accuracy of the direct 
inversion method reconstructions, because it damps higher order modes. The
input and output (reconstructed) coefficients $a_{\ell \ell}$ for the multipoles $\ell=2,\dots,5$
are compared in \fig(\ref{alm_corr}) and (\ref{alm_corr2}). Only those modes with 
$m=\ell$ are shown because, as will be shown in \sect\ref{cov_mat}, they (and the corresponding
 modes with $m=-\ell$) have the largest reconstruction uncertainties as a result of the removal 
of a large fraction of their power by the cut. Thus, they are the most sensitive to the method used 
in their estimation. Nevertheless, comparison of the other modes gives qualitatively similar
results.
 
\fig(\ref{alm_corr}) shows that the PE filter recovers the quadrupole and octopole amplitudes
with a similar accuracy to the direct inversion method. However, for
higher order modes, the PE filter performs better. This does not appear to be connected to
the use of $\ell_{\rm{rec}}=10$ for the direct inversion method; increasing this value
to $\ell_{\rm{rec}}=30$ yields even worse reconstructions (\fig(\ref{alm_corr2})). 
The higher accuracy of the PE filter reconstructions is a consequence of 
explicitly accounting for the aliasing of higher order multipoles into the lower
multipole range in the filter construction.


\section{Application to the \emph{WMAP} data} \label{application}
The PE filter was applied to \emph{WMAP} maps with 
$\ell_{\rm{max}}=30$ and $\ell_{\rm{rec}}=10$. We are mostly interested
in the analysis of the lowest multipoles $\ell=2,\dots,5$ because these 
are the most sensitive to the cut and foreground emission. 
The monopole and dipole were not restored because they do not provide any 
relevant cosmological information.
In order to construct the filter matrix, the \emph{WMAP} best-fitting $\Lambda$CDM model 
($\Omega_\Lambda = 0.73$, $\Omega_m=0.27$) with a running primordial spectral 
was used. This choice is based on the fact that the
spectrum provides a better fit to the \emph{WMAP} data on large angular scales compared to a simple power-law 
spectrum. However, tests with a power-law spectrum yields similar results.
The noise covariance matrices were constructed in the usual fashion, with pixel variances determined
by the number of observations $N_{\rm{obs}}(p)$ in pixel $p$ as provided by the \emph{WMAP} team. 

The power spectrum will provide a useful complement to the studies of the
multipole maps. 
We will use an estimator of the power spectrum given by 
\begin{equation} \label{d} 
\hat{C}_\ell=\frac{\sum_{m=-\ell}^{\ell} |\hat{a}_{\ell m}|^2}{2 \ell+1} \ , 
\end{equation}
where $\hat{a}_{\ell m}$ denotes the restored harmonic
coefficients. The definition of the PE filter ensures that the estimator is unbiased. Given the error 
covariance matrix $\mathbf{M}$, one can estimate the uncertainty
in the quantity $\hat{C}_\ell$ by
\begin{equation} \label{cl_err} 
\sigma_\ell^2=\frac{2}{(2 \ell+1)^2} \sum_{m,m' =-\ell}^\ell 
\mathbf{M}_{ij}\left(2 \hat{a}_i \hat{a}_j -\mathbf{M}_{ij}\right) \ ,
\end{equation} 
where the indices $i,j$ in $\ell$-ordering are expressed by
\mbox{$i=\ell^2+\ell+m+1$}. This estimation of the uncertainty 
takes into account only errors induced by the sky cut and noise;
it does \emph{not} take into account cosmic variance related errors which enter through the assumption of a 
cosmological power spectrum to build the PE filter.
\tab(\ref{tab_clerr_r}) indicates the reconstruction accuracy of the method.
 
\begin{table}
\begin{center}
\caption{
	 Rms difference between the true and estimated power spectra 
	 $C_\ell^{\rm{error}}$ (in $\mu K^2$) and the correlation coefficients
	 $r_{\ell \ell}$ of the $\ell$-order, $m=\ell$ modes estimated 
	 from a set of 10 000 simulations of the \emph{WMAP} data analysed after 
         application of the Kp2 mask. In the third column, the signal to error ratio is
	 shown where $\hat{C}_\ell$ is the estimated power spectrum of the
	 \emph{WMAP} multipoles maps given by PE filtering.}    
\label{tab_clerr_r}  
{\scriptsize 
\begin{tabular}{cccc} 
\hline 
$\ell$ & $C_\ell^{\rm{error}}$  & $\hat{C}_\ell^{}/C_\ell^{\rm{error}}$ & $r_{\ell \ell}$ \\
\hline		
2& 69& 2.8& 0.99\\ 
3& 41& 12& 0.98\\ 
4& 26& 9.6& 0.97\\ 
5& 19& 17& 0.96\\ 
6& 14& 5.2& 0.94\\ 
7& 12& 11& 0.92\\ 
8& 9.0& 6.3& 0.90\\ 
9& 7.6& 7.6& 0.87\\ 
10& 6.2& 7.3& 0.84\\
\hline		
\end{tabular}}
\end{center}
\end{table} 


\subsection{The multipoles maps} \label{maps}
\begin{figure*}
\begin{center}
\includegraphics[angle=0.,width=0.7\textwidth]{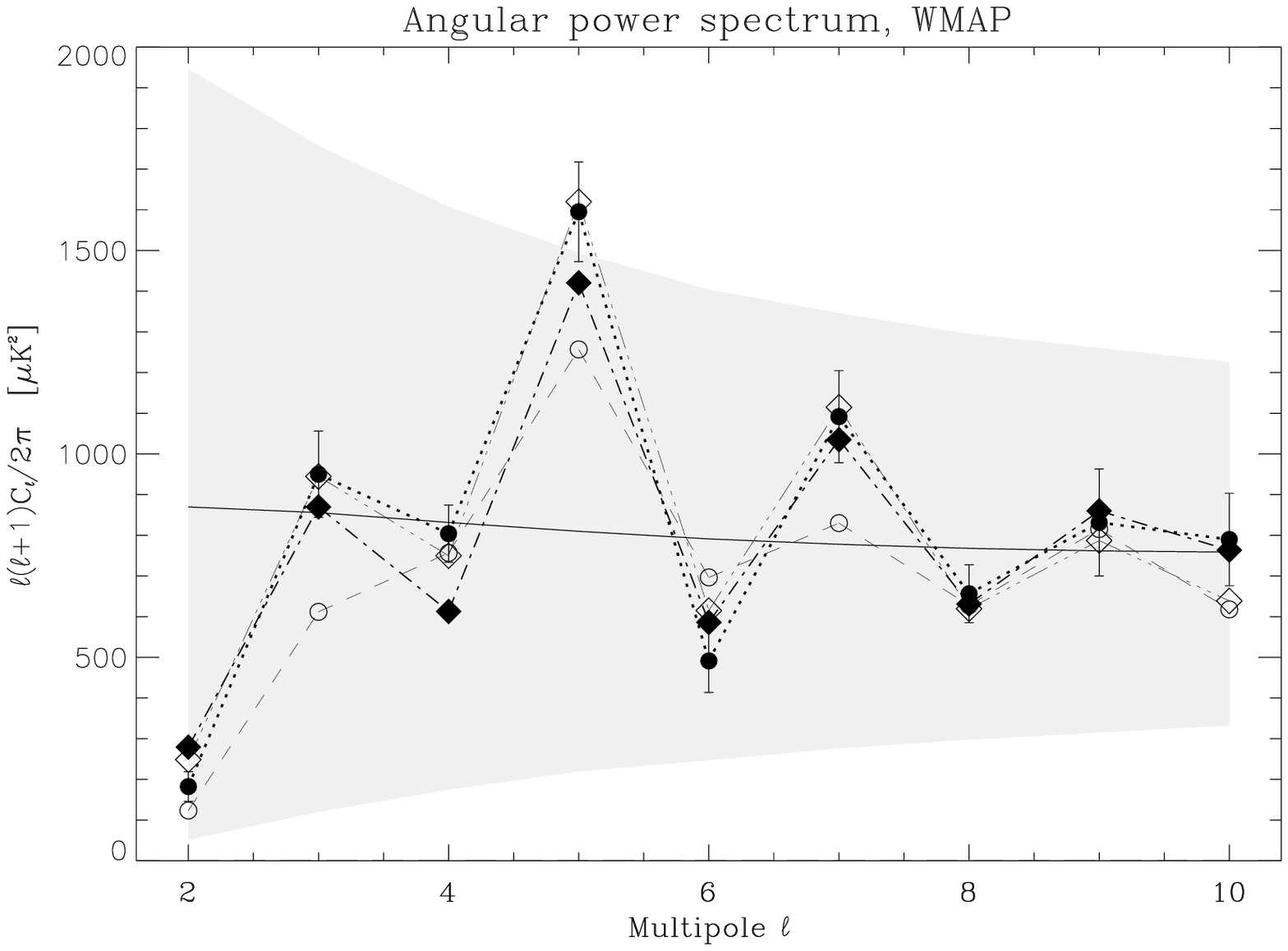}
\caption{Angular power spectrum of the \emph{WMAP} data.
The power spectrum from the Hinshaw et al. (2003) cut sky \emph{WMAP} analysis (open circles) is compared
with those obtained by the PE filtering of the TOH
(filled diamonds), ILC (open diamonds) and   
corrected co-added map (filled circles), masked by the
Kp2 mask. The error bars ($68\%$ confidence level) of our \emph{WMAP} 
power spectrum estimates are shown. They do not take into account cosmic
variance. The major contribution to the errors comes from the correlations
between the restored multipoles and between the restored and higher multipoles. The
noise contribution is negligible. The smooth curve shows the \emph{WMAP} team
best-fitting $\Lambda$CDM model ($\Omega_\Lambda=0.73$ and $\Omega_m=0.27$)
with running spectral index. The grey band indicates the cosmic variance
errors (95 per cent confidence level).} \label{cl_wmap} 
\end{center}
\end{figure*}

Maps of the recovered multipoles computed from the \emph{WMAP} corrected co-added
map are compared to the multipoles of the cleaned TOH and 
ILC maps in \fig (\ref{maps_wmap}). 
Because the details of the noise model are essentially negligible for the estimation of 
low-order multipoles, 
we have also used the PE filter as computed for the corrected co-added map 
to analyse the TOH and ILC maps.
The comparison confirms the previous results of estimations of the
multipole amplitudes. Small discrepancies are seen only at $\ell=2$, where the hot spots 
are closer to the Galactic equator and the amplitude is smaller for our results.
The power spectrum of the maps is shown in \fig(\ref{cl_wmap}) and summarised in 
\tab(\ref{tab_dt}). 
The cut sky analysis of \citet{hinshaw} shows substantially less power
at $\ell=3,5$ and 7 than for the other estimates. \citet{efstathiou}
has claimed that this is a consequence of the quadratic nature of this estimator.
Our linear estimator results, and in particular the ILC results in \tab(\ref{tab_dt}), 
are in good agreement with those of \citet{efstathiou}.

\subsection{Dependence on the assumed angular power spectrum}
\label{higher_modes}

\begin{table*}
\begin{center}
\caption{
	 Power spectrum amplitudes $\Delta T_\ell^2=\ell (\ell+1)
	 C_\ell/2\pi$ (in $\mu K^2$) for the foreground corrected 
         \emph{WMAP} frequency maps and related combinations estimated by application of the PE filter
	 for the Kp2 sky coverage. \emph{Q+V+W} denotes the corrected co-added map
	 and $^\ast$ denotes estimation on the full sky. The direct
	 inversion method was applied to the corrected co-added map for the
	 Kp2 mask. Uncertainties do
	 not take into account cosmic variance.} 
\label{tab_dt}
{\scriptsize
\begin{tabular}{cccccccccc}
\hline
 &$\Delta T^2_2$  & $\Delta T^2_3$ & $\Delta T^2_4$  & $\Delta T^2_5$ &
 $\Delta T^2_6$ & $\Delta T^2_7$ & $\Delta T^2_8$ & $\Delta T^2_9$ & $\Delta T^2_{10}$ \\
\hline		
\emph{K} & 963 $\pm$ 79 & 1710 $\pm$ 160 & 3290 $\pm$ 220 & 2070 $\pm$ 130 & 710
 $\pm$ 110 & 1060 $\pm$ 120 & 1320 $\pm$ 120 & 1330 $\pm$ 150 
 & 1680 $\pm$ 180 \\
\emph{Ka} & 176 $\pm$ 26 & 1060 $\pm$ 120 & 1190 $\pm$ 110 & 1550 $\pm$ 110 & 312
 $\pm$ 55 & 960 $\pm$ 100 & 786 $\pm$ 84 & 920 $\pm$ 130 & 1100 $\pm$ 140  \\
\emph{Q} & 158 $\pm$ 32  & 990 $\pm$ 110  & 875 $\pm$ 80  & 1610 $\pm$ 120 & 459
 $\pm$ 77 & 1100 $\pm$ 110 & 683 $\pm$ 75 & 880 $\pm$ 130 & 840 $\pm$ 120 \\
\emph{V} & 206 $\pm$ 40  & 930 $\pm$ 100  & 755 $\pm$ 63  & 1600 $\pm$ 120 & 523
 $\pm$ 79 & 1070 $\pm$ 110 & 644 $\pm$ 69 & 800 $\pm$ 130 & 750 $\pm$ 110 \\
\emph{W} & 219 $\pm$ 42  & 910 $\pm$ 100  & 804 $\pm$ 67 & 1580 $\pm$ 130 & 553
 $\pm$ 79 & 1110 $\pm$ 120 & 645 $\pm$ 69 & 780 $\pm$ 130 & 770 $\pm$ 110 \\
\emph{Q+V+W} & 182 $\pm$ 36  & 950 $\pm$ 110  & 804 $\pm$ 70  & 1600 $\pm$ 120 &
 491 $\pm$ 78 & 1090 $\pm$ 110 & 656 $\pm$ 71 & 830 $\pm$ 130 & 790 $\pm$
 110 \\
Direct inversion & 322 $\pm$ 36 & 873 $\pm$ 96 & 1530 $\pm$ 120 & 1370 $\pm$ 110 & 569 $\pm$ 78
 & 870 $\pm$ 100 & 1037 $\pm$ 99 & 950 $\pm$ 140 & 1090 $\pm$ 130 \\
ILC & 248 $\pm$ 45  & 950 $\pm$ 110  & 750 $\pm$ 61  & 1620 $\pm$ 130 & 615
 $\pm$ 85 & 1120 $\pm$ 120 & 619 $\pm$ 66 & 790 $\pm$ 130 & 639 $\pm$ 98 \\
TOH & 279 $\pm$ 49  & 870 $\pm$ 98 & 613 $\pm$ 49  & 1420 $\pm$ 110 & 586
 $\pm$ 89 & 1040 $\pm$ 110 & 631 $\pm$ 63 & 860 $\pm$ 140 & 760 $\pm$ 110 \\
ILC$^\ast$ & 195 $\pm$ 40  & 1050 $\pm$ 110  & 834 $\pm$ 66  & 1670 $\pm$
 130 & 606 $\pm$ 81 & 1290 $\pm$ 140 & 660 $\pm$ 68 & 700 $\pm$ 130 & 780
 $\pm$ 110 \\
TOH$^\ast$ & 202 $\pm$ 41  & 866 $\pm$ 99  & 651 $\pm$ 50  & 1350 $\pm$ 110
 & 608 $\pm$ 90 & 1010 $\pm$ 110 & 681 $\pm$ 72 & 710 $\pm$ 120 & 890 $\pm$
 120 \\
\hline
\end{tabular}}
\end{center}
\end{table*}

\begin{figure*}
\begin{center}
\includegraphics[angle=0.,width=0.7\textwidth]{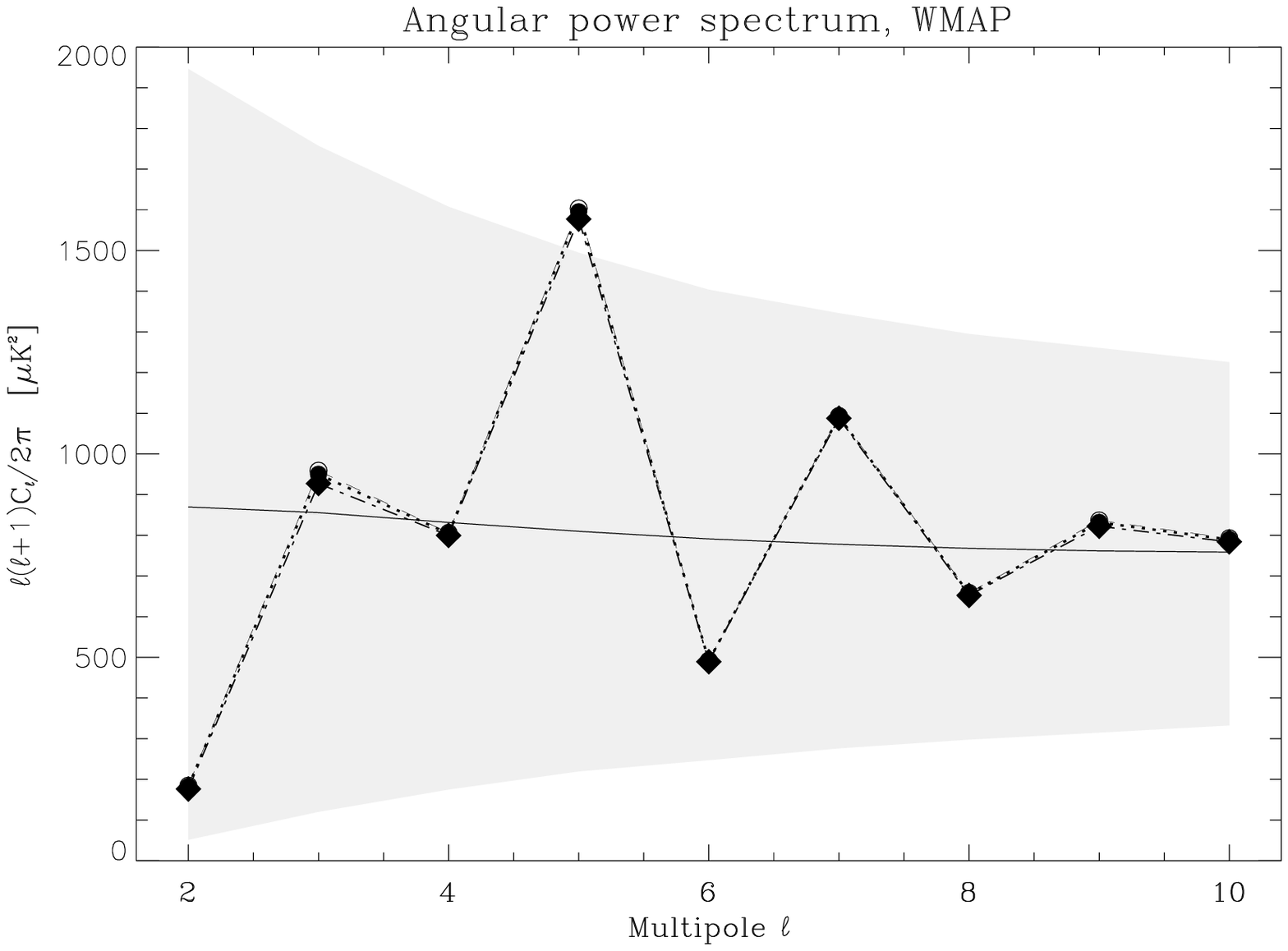}
\caption{Sensitivity of the derived \emph{WMAP} angular power spectrum
to the input power spectrum used in the construction of the PE
filter. Filled circles correspond to the spectrum of the multipoles obtained assuming
the best fit to the \emph{WMAP} data $\Lambda$CDM model power spectrum (indicated
by the solid line). The open circles and filled
diamonds correspond, respectively, to the same assumed power spectrum
but after adding and subtracting the square root of the cosmic variance
(indicated by the shaded band, 95 per cent confidence level).} \label{cl_wmap_w+-}
\end{center}
\end{figure*}

Because construction of the filters requires some assumption about the CMB
anisotropy power spectrum $C_\ell$, one should investigate how sensitive the amplitudes of the 
restored multipoles are to this choice and thus quantify the uncertainties in the amplitudes
related to the assumption. To answer these questions we have applied  
filters constructed using two other power spectra: the canonical one described above 
but with the square root of the cosmic variance either added or subtracted. 
In \fig(\ref{cl_wmap_w+-}), we see that  the power spectrum of the resultant maps
is changed only weakly by these perturbations, up to few percent. 
For the \emph{WMAP} analysis presented here, the PE filtering method gives robust
estimates of the amplitudes of the low-order multipoles, essentially independent of the 
assumed input power spectrum.

\subsection{Dependence on frequency} \label{depen_freq}
Comparison of the multipoles maps determined from the foreground corrected
\emph{WMAP} data frequency-by-frequency allows us to determine the extent to which the
derived structures are cosmological and the degree of residual foreground contamination.
In \fig(\ref{maps_wmap_freq}), it is seen that the foreground
cleaning of the \emph{K}-band map gives the worst results. The multipole maps
differ substantially from the rest of the maps showing excess of the power near
the Galactic plane. Nevertheless, apart from the quadrupole, there is 
a qualitative similarity in the distribution of hot and cold spots relative to the other frequency channels.
In particular, the octopole is planar with approximately the same preferred axis, along which the
power is suppressed, for all frequencies. The quadrupole is also planar, although changes of the plane are more
pronounced, especially for the \emph{K} and \emph{Ka} bands. Small deviations between
the \emph{Q}, \emph{V} and \emph{W} bands indicates that, after foreground cleaning, these maps are dominated by the CMB
anisotropy component and can be treated as a reliable picture of the CMB multipoles.

\begin{figure*}
\begin{center}
\includegraphics[angle=0.,width=0.7\textwidth]{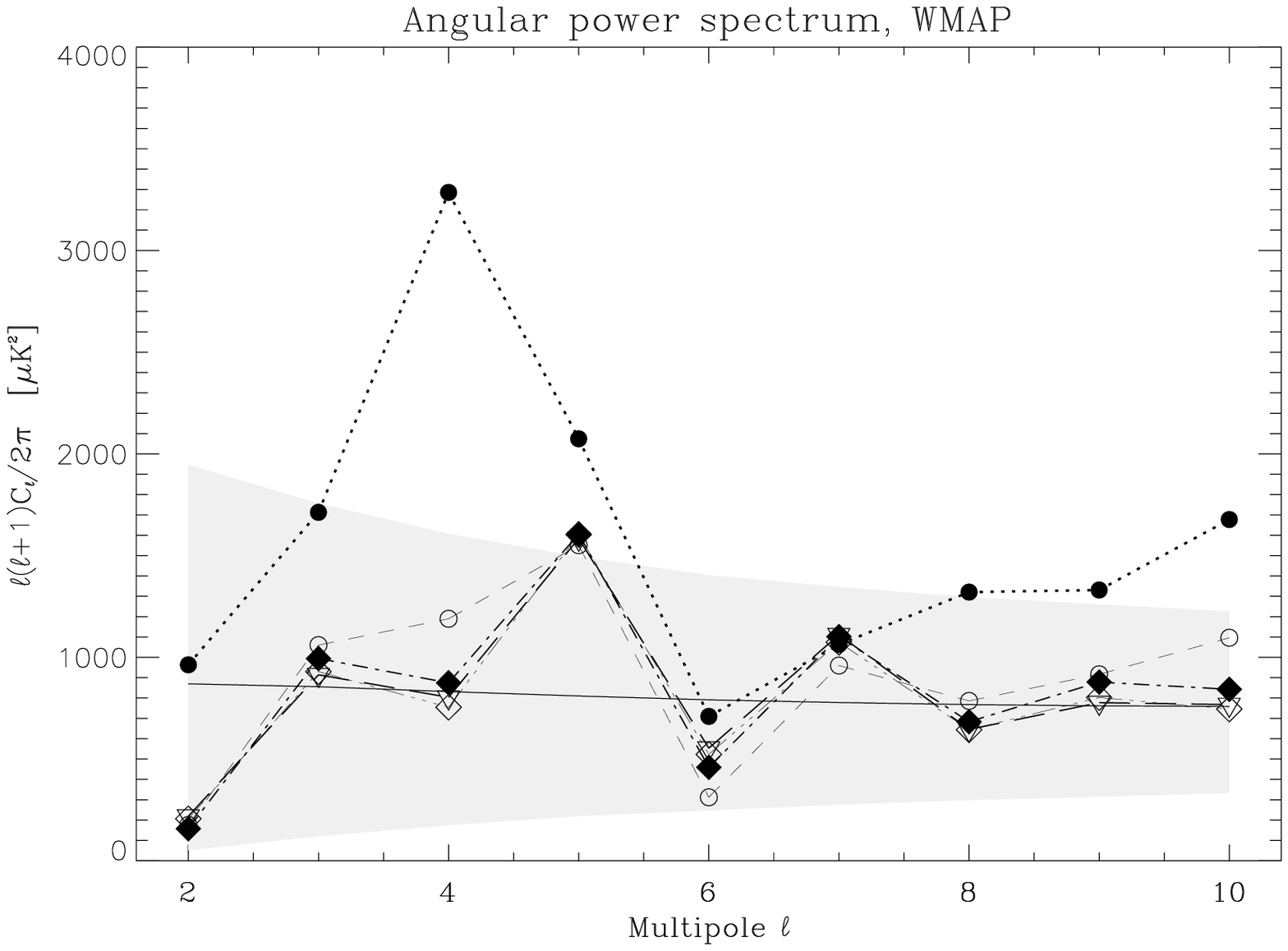}
\caption{The angular power spectrum of the \emph{WMAP} data
for \emph{K}-band (filled circles), \emph{Ka}-band (open circles), \emph{Q}-Band (filled diamonds), 
\emph{V}-Band (open diamonds) and
\emph{W}-Band (triangle) maps obtained by PE filtering for the Kp2 mask. The smooth curve shows the \emph{WMAP} team
best-fitting $\Lambda$CDM model ($\Omega_\Lambda=0.73$ and $\Omega_m=0.27$)
with running spectral index. The grey band indicates the cosmic variance
errors (95 per cent confidence level).} \label{cl_wmap_kqvw}
\end{center}
\end{figure*}

The corresponding power spectra shown in \fig(\ref{cl_wmap_kqvw}) and summarised in 
\tab(\ref{tab_dt}) clearly show that the applied foreground correction does not remove 
all the contamination for the \emph{K} and (partially) \emph{Ka} bands. This suggests that the Kp2 cut is not
sufficient to remove the strong signals near to the Galactic plane at those frequencies
(despite the fact that the Kp2 mask is derived from considerations of the thresholded
temperature distribution of the \emph{K}-band map). The strong synchrotron and free-free emission extends to
higher Galactic latitudes at these lower frequencies.

\subsection{Dependence on the sky cut} \label{depen_mask} 
More substantial changes of the multipoles are seen with then application of more aggressive sky 
cuts. In \fig(\ref{maps_wmap_cut}), we show the multipoles determined from the \emph{WMAP}
corrected co-added data after masking by the Kp2, Kp0, 20+ and 30
masks. All multipoles lose their characteristic features for
mask 30. Nevertheless, up to sky cut 20+ the multipoles $\ell=3$ and
$\ell=5$ do not change significantly. The multipoles $\ell=2$ and $\ell=4$ are 
more sensitive to the mask and change substantially even for smaller
cuts. 

\begin{figure*}
\begin{center}
\includegraphics[angle=0.,width=0.7\textwidth]{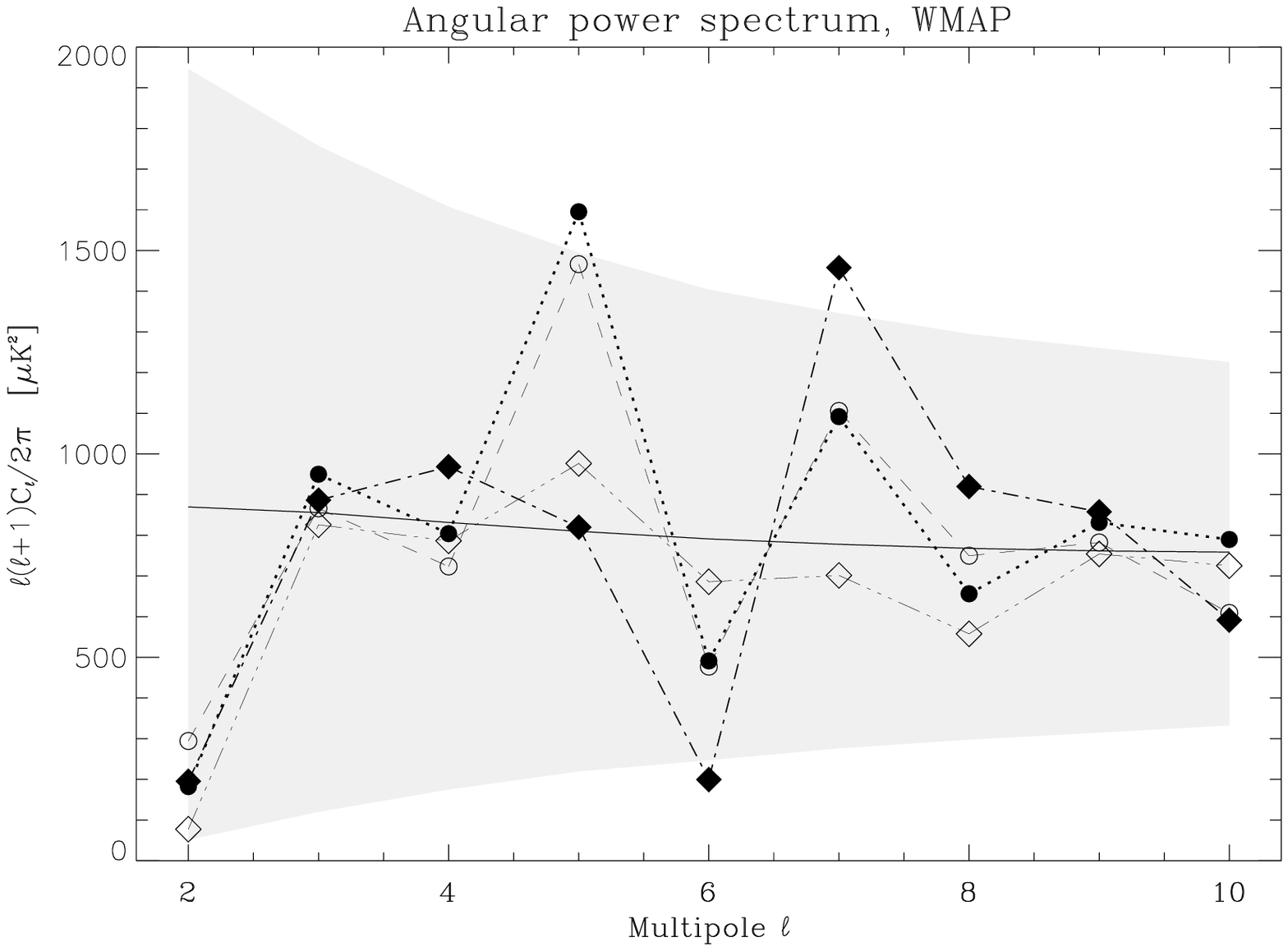}
\caption{The angular power spectra of the \emph{WMAP} data
obtained by the PE filtering of the corrected co-added \emph{Q+V+W} map after
masking with the Kp2 (filled circles), Kp0 (open circles), 20+
(filled diamonds) and 30 (open diamonds) masks. The smooth curve shows the \emph{WMAP} team 
best-fitting $\Lambda$CDM model ($\Omega_\Lambda=0.73$ and $\Omega_m=0.27$)
with running spectral index. The grey band indicates the cosmic variance
errors (95 per cent confidence level).} \label{cl_wmap_cut}
\end{center}
\end{figure*}

The corresponding power spectra are presented in \fig(\ref{cl_wmap_cut}) and
\tab(\ref{tab_dt_mask}). It should be noticed that generally the uncertainties increase
rapidly with the width of the sky cuts. Because our estimation of the errors (\ref{cl_err})
depends on the multipole power, the few exceptions from this rule, 
$\ell=6$ for the 20+ mask, $\ell=7,8,9$ for the 30 mask, are a consequence of the substantially
lower amplitudes of the multipoles estimated in those cases. 
Notice that for very low power, the estimation of the square of the error can be negative. 
Because we have such a situation for the quadrupole estimated for the cut 30, we do not quote
the corresponding uncertainties in the table.  

\begin{figure*}  
\begin{center} 
\includegraphics[angle=0.,width=1.\textwidth]{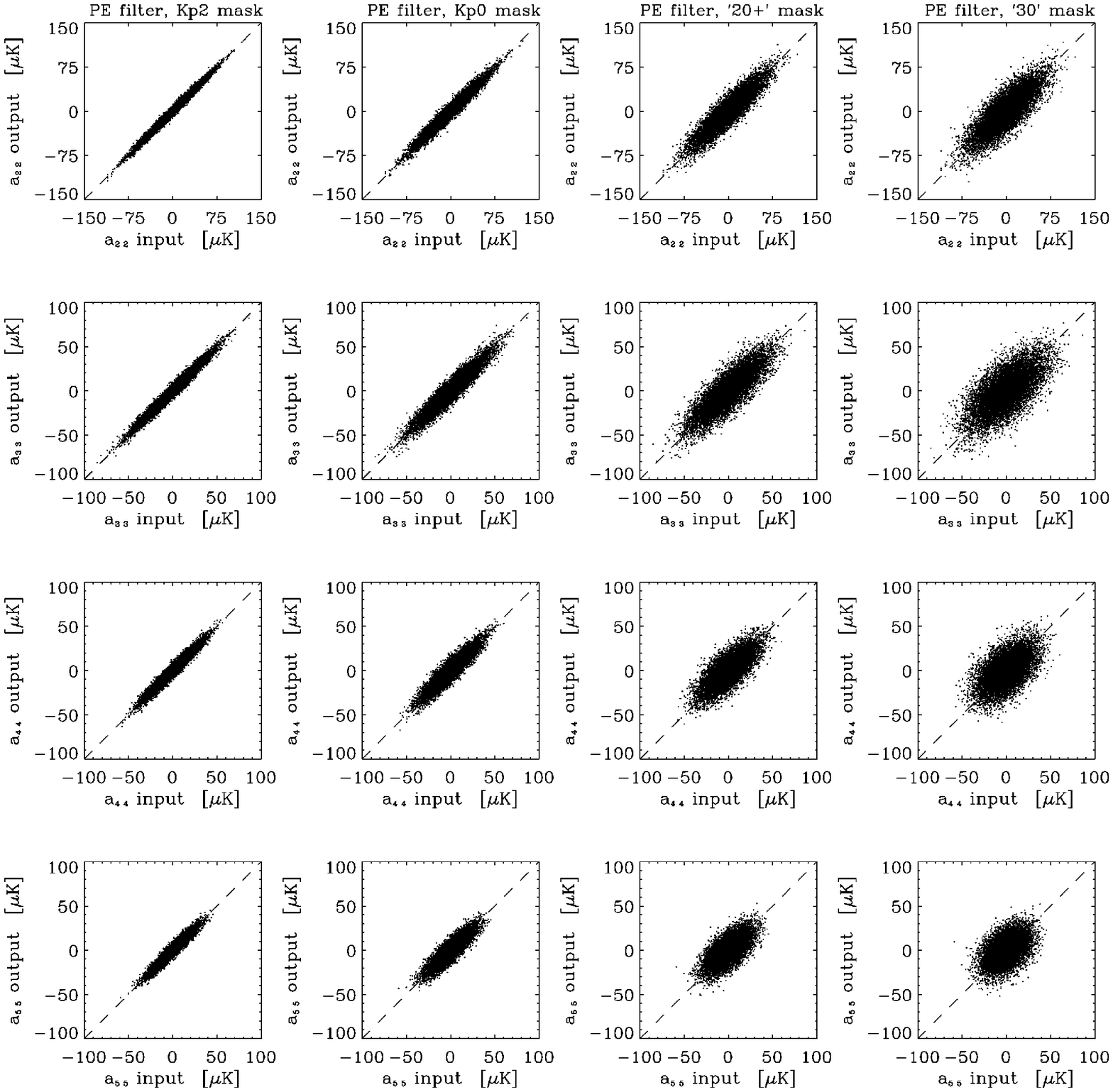}
\caption{Comparison of the accuracy of estimation of the $a_{\ell,\ell}$
modes using the PE filtering reconstruction method. 
The abscissae give the input values of the harmonic coefficients used to generate the simulated skies. 
The ordinates give the output values determined after application of the Kp2 mask (first column), 
Kp0 mask (second column), 20+ mask (third column) and 30 mask (fourth column).} \label{alm_corr_k2_k0}
\end{center}   
\end{figure*}

The general trend for increasing uncertainties with increasing sky cut is easily
seen in \fig(\ref{alm_corr_k2_k0}) where
linear regressions, analogous to that in
\sect\ref{comparision}, is shown for various masks. It should be noticed
that uncertainties in the estimation of the multipoles $\ell=4, 5$  is especially large for
the cut 30. The true and reconstructed coefficients are nearly
uncorrelated. Slightly better correlations are seen for modes with $m\neq
-\ell,\ell$, though the trend is still maintained. The poor 
correlation between input and recovered coefficients implies that for $\ell>5$
their estimation is no longer credible. 
Indeed, the large uncertainties associated with the widest cut 30 helps to explain why the appearance
of the estimated \emph{WMAP} multipoles is different from that
estimated with narrower masks. Nevertheless, the multipole amplitudes remain statistically consistent
over the range of masks at approximately the $2\sigma$ level.

\begin{table}
\begin{center}
\caption{
	 Power spectra of the multipoles $\Delta T_\ell^2=\ell (\ell+1)
	 C_\ell/2\pi$ (in $\mu K^2$) estimated by the PE filtering after applying
	 various masks. Uncertainties do not take into account cosmic variance.} 
\label{tab_dt_mask}
{\footnotesize
\begin{tabular}{cccccc}
\hline
 & & \multicolumn{4}{c}{Mask} \\
 & & Kp2 & Kp0 & 20+ & 30  \\
\hline		
 & $\ell=2$ & 182 $\pm$ 36 & 294 $\pm$ 88 & 200 $\pm$ 130 & 77 \\
 & $\ell=3$ & 950 $\pm$ 110 & 870 $\pm$ 170 & 890 $\pm$ 310 & 830 $\pm$ 380 \\
 & $\ell=4$ & 804 $\pm$ 70 & 723 $\pm$ 94 & 970 $\pm$ 280 & 790 $\pm$ 310 \\
 & $\ell=5$ & 1600 $\pm$ 120 & 1470 $\pm$ 170 & 820 $\pm$ 280 & 980 $\pm$ 450 \\
$\Delta T_\ell^2$ & $\ell=6$ & 491 $\pm$ 78 & 480 $\pm$ 110 & 200 $\pm$ 54 & 690 $\pm$ 260 \\
 & $\ell=7$ & 1090 $\pm$ 110 & 1110 $\pm$ 190 & 1460 $\pm$ 380 & 700 $\pm$ 240 \\
 & $\ell=8$ & 656 $\pm$ 71 & 750 $\pm$ 130 & 920 $\pm$ 250 & 560 $\pm$ 130 \\
 & $\ell=9$ & 830 $\pm$ 130 & 780 $\pm$ 180 & 860 $\pm$ 250 & 750 $\pm$ 220 \\
 & $\ell=10$ & 790 $\pm$ 110 & 610 $\pm$ 140 & 590 $\pm$ 140 & 730 $\pm$ 170 \\
\hline
\end{tabular}}
\end{center}  
\end{table}


\section{Uncertainties of the multipoles maps} \label{errors}
One of the merits of the PE filtering method is the possibility to directly estimate 
the uncertainties in the derived results, and the preservation of the well-defined 
noise properties of the original maps. This is a clear advantage of the method over 
the MEM applied by the \emph{WMAP} team
\citep{bennet4}
which results in anisotropy maps with computationally 
complex noise properties which are 
largely inappropriate for most CMB analyses. 
Indeed, it is generally true that the noise evaluation for linear
methods are computationally more feasible than for non-linear methods
such as MEM. Here we derive errors both in pixel and spherical harmonic space.

\subsection{The maps of the standard deviation} \label{stan_dev} 
To study the spatial distribution of the uncertainties of the restored
multipole maps, we use the simulations of the \emph{WMAP} corrected
co-added map as described in \sect\ref{lmaxlrec}. The map
\begin{equation} \label{st_dev}
\Delta^2(p)_{\rm{error}}=\sum_{\ell_1,m_1} \sum_{\ell_2,m_2}
\left<a_{\ell_1,m_1}^{\rm{error}} a_{\ell_2,m_2}^{\rm{error}}\right> Y_{\ell_1,m_1}(p)
Y_{\ell_2,m_2}(p) \ ,
\end{equation} 
shows the spatial distribution of the squared standard deviation of the
difference between the restored and true multipole maps. 
$a_{\ell_1,m_1}^{\rm{error}}$ are the corresponding decomposition coefficients
of the difference maps and terms $\left<a_{\ell_1,m_1}^{\rm{error}} 
a_{\ell_2,m_2}^{\rm{error}}\right>$ are averaged over 10 000
simulations. Using the identity
\begin{eqnarray} \label{e}
Y_{\ell_1,m_1}(p) Y_{\ell_2,m_2}(p)&=&\sum_{\ell,m} 
\sqrt{\frac{(2\ell_1+1)(2\ell_2+1)(2\ell+1)}{4\pi}} \nonumber \\
& & \times \left(
\begin{array}{ccc}
\ell_1 & \ell_2 & \ell \\
m_1 & m_2 & m
\end{array} \right)
 \left(\begin{array}{ccc}
\ell_1 & \ell_2 & \ell \\
0 & 0 & 0
\end{array} \right)
Y_{\ell,m}(p)\ , \nonumber
\end{eqnarray}
where $\left(\begin{array}{ccc}
\ell_1 & \ell_2 & \ell \\
m_1 & m_2 & m
\end{array} \right) $ is the Wigner 3j symbol,
the expression (\ref{st_dev}) can be written as $$\Delta^2(p)_{\rm{error}}=\sum_{\ell,m}
b_{\ell,m} Y_{\ell,m}(p) \ ,$$ where $b_{\ell,m}$ are the appropriate coefficients.
The $\Delta_{\rm{error}}^2$ map is composed of modes $0 \le \ell \le 2\ell_1$,
because $\ell$ obeys the triangular
condition ($\ell=\ell_1+\ell_2,\ell_1+\ell_2-1,\dots,|\ell_1-\ell_2|$) and
$\ell_1=\ell_2$. 
\\

The $\Delta_{\rm{error}}^2$ maps computed for the Kp2 mask  
are shown in \fig(\ref{map_st_dev_wmap}) in logarithmic scale. As expected, as a result of
the lack of the data in the cut part of the sky the  
reconstructed multipoles have the largest uncertainties in this region, especially
close to the Orion and Ophiuchus exclusion zones. The multipoles $\ell=2$ and $3$
have the smallest uncertainties inside the cut. This is a consequence of 
the fact that the multipole maps characterize the CMB anisotropy on angular scales
$\sim 1/\ell$.  
If this is bigger than the angular width, $\Delta \theta$, of the mask
then the excised region of the multipole map can be easily extrapolated from the data
outside of the cut. Otherwise, there will be fluctuations with typical angular size smaller
than $\Delta \theta$, which will be completely contained within the cut and
their reconstruction from the data will be much less accurate. The
statistical isotropy assumption used in our approach is then very helpful
to reduce uncertainties in these cases. For the Kp2 mask, the
widest part, cutout toward the Galactic centre, has width $\Delta \theta
\approx 40^\circ$. 

The $\Delta_{\rm{error}}^2$ map corresponding to the sum of the multipoles 
$\ell=2,\dots,5$ shows much bigger uncertainties than for the corresponding 
individual multipoles.  
This is a result of the additional contributions from the correlations 
between errors in the multipoles. These correlations are easily seen in the covariance 
matrix $\mathbf{M}$.

\subsection{The covariance matrix} \label{cov_mat}
The map of the standard deviation does not show all of the information about the
errors. The covariance matrix $\mathbf{M}$ shows that the uncertainties are correlated in the
spherical harmonic space.  

\begin{figure*}
\begin{center}
\includegraphics[angle=0.,width=0.8\textwidth]{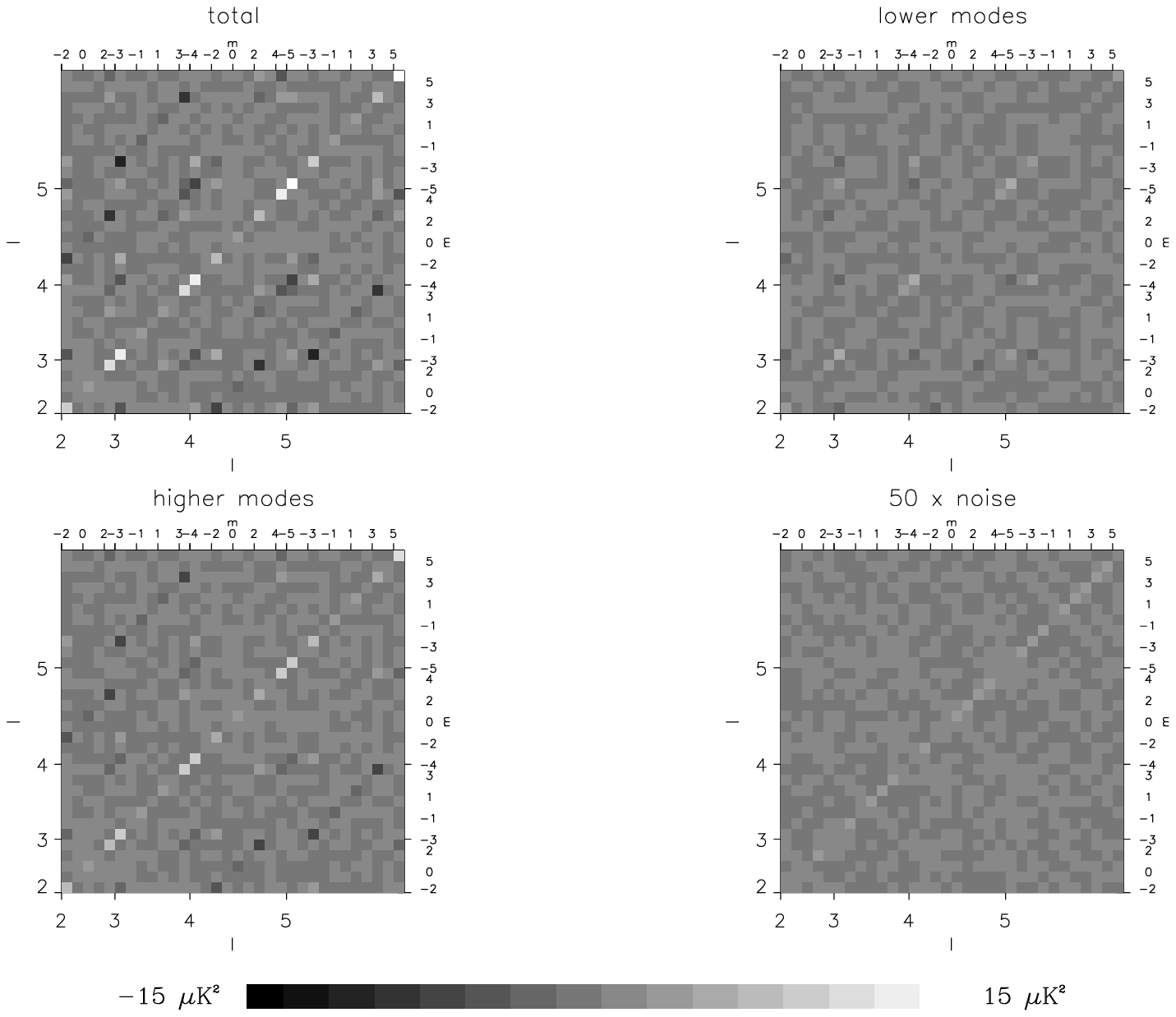}
\caption{Decomposition of the covariance matrix for the PE filtering
of the \emph{WMAP} data into specific contributions is shown. The total covariance matrix
(upper left figure) is compared with the components induced by:
correlations between restored multipoles (upper right figure), 
correlations of restored multipoles with higher order multipoles (bottom left figure) 
and noise correlations (bottom right figure). Only that part of
the matrices corresponding to the range $\ell=2,\dots,5$ are shown. The
noise component is multiplied by 50 to render it visible. Maximal correlation
is at most $\sim 14\ \mu \rm{K}^2$ and minimal down to $\sim -11\ \mu \rm{K}^2$. The
modes are ordered such that index $i$ is given by $i=\ell^2+\ell+m+1$.} \label{cov_noise_wmap_w}
\end{center}
\end{figure*}

As mentioned in \sect\ref{problem}, the covariance matrix, 
$\mathbf{M}$, consists of
three components: the noise component and, induced purely by the cut,
contributions from correlations between the restored multipoles and correlations
with higher order multipoles.
In \fig(\ref{cov_noise_wmap_w}) we show that
part of the covariance matrix corresponding to 
the multipoles $\ell=2,\dots,5$ for the PE filtering of the \emph{WMAP}
corrected co-added map, together with the corresponding sub-matrices for the
separate components.
The largest contribution arises from correlations with higher
order modes. The correlations between the low-order reconstructed multipoles 
are weaker although still non-negligible. The smallest contribution comes
from the noise component, which is plotted multiplied by a factor of 50 to render it visible.
The approximately axisymmetric cut causes stronger correlations between
modes with $\ell$ value different by 2.
Notice that the modes with the biggest reconstruction uncertainties are for
$m=-\ell$ and $m=\ell$, which is a consequence of removing a
large fraction of the power of these modes, mainly concentrated in the
Galactic plane, by the cut.   


\section{Foreground separation} \label{foreground}
In this section, we will reconsider the sensitivity of the estimated
CMB multipoles to the Galactic foregrounds. 
To achieve this, we derive a CMB anisotropy map by the spectral 
decomposition of multi-frequency
\emph{WMAP} data by making assumptions about the frequency dependence
of the foreground components.  
The advantage of this method is that,
unlike with the template fits of
\citet{bennet4}, we assume nothing about the morphological distribution of the
foreground emission. Thus foregrounds which are not well traced by the
existing templates, but with similar spectra, e.g. the free-free haze
proposed by \citet{finkbeiner}, are taken into account. The
disadvantage is that, to separate the components properly, we need to
know their spectral dependence. 
Given that our intention is not to provide a definitive 
foreground separation analysis, 
we have considered a broad range of spectral index values
to aggressively test the dependence of the
reconstructed low-order multipoles on any foreground assumptions.
Nevertheless, it should be recognised
that our assumptions may lead to \emph{worse} foreground subtraction
than in the \emph{WMAP} analysis, although this is unlikely given that
the range of spectral properties adopted here 
are well motivated by previous results in the literature.

We consider a method following \citet{bennet2} and
the formalism from \citet{dodelson}.
We suppose now that we observe the sky in $m$ different frequency channels,
so that at each point $\mathbf{r}$ on the sky we measure $m$ different
temperatures $d_i(\mathbf{r})$, $i=1,\dots,m$. Let us assume also that we know
the temperature spectra $f_j(\nu)$ of all $n$ components $x_j(\mathbf{r})$,
$j=1,\dots,n$. Then we can write
\begin{equation} \label{multifreq}
\mathbf{d}(\mathbf{r})=\mathbf{F}\cdot \mathbf{x}(\mathbf{r})+\mathbf{n}(\mathbf{r}) \ ,
\end{equation}
where the vector $\mathbf{n}(\mathbf{r})$ corresponds to the instrumental noise
in the various channels and $\mathbf{F}$ (not to be confused with the filter
matrix used in \sect\ref{sec:filters}) is the $m\times n$ frequency response
matrix given by 
\begin{equation}
F_{ij}\equiv \int_0^\infty w_i(\nu) f_j(\nu) d\nu \ ,
\end{equation}
$w_i(\nu)$ being the frequency response (bandwidth) of the $i$th channel. Here, we will
approximate $w_i(\nu)$ by Dirac's delta $w_i(\nu)=\delta(\nu-\nu_i)$ where
$\nu_i$ corresponds to the mean frequency in the $i$th channel, so
that $F_{ij}=f_j(\nu_i)$.
If the number of frequencies, $m$, equals the number of components $n$ and
noise is negligible, the separation of the components can be obtained by inverting
the frequency response matrix, 
\begin{equation}
\mathbf{x}(\mathbf{r})=\mathbf{F}^{-1} \cdot \mathbf{d}(\mathbf{r}). 
\end{equation}
For the CMB component, if temperature is expressed in thermodynamic
units, $F_{i1}=1$ over all frequencies as a result of the blackbody nature of the spectrum.
For the range of frequencies of interest here, 
the foreground spectra may be reasonably well modelled by a power law in antenna
temperature, $f_j(\nu) \propto
\nu^{\beta_j}$, where the spectral index $\beta_j$ depends on the component. 
To express the spectral dependence in thermodynamic units, 
the function $f_j(\nu)$ must include the frequency dependent antenna to 
thermodynamic temperature conversion factor, $p(\nu)$. Thus, 
finally the functions take
the form $f_j(\nu)=p(\nu) \left(\nu/\nu_{0,j}\right)^{\beta_j}$,
where $\nu_{0,j}$ is the reference frequency for component $j$ at which the
foreground amplitude is normalised.

For the \emph{WMAP} data, sky maps are available at 5 frequencies: \emph{K}, \emph{Ka},
\emph{Q}, \emph{V}, and \emph{W} bands\footnote{Note that, before proceeding with the separation, these
were smoothed to a common angular resolution of $1^\circ$.}
($i=1,\dots,5$, respectively) and these may be
interpreted as consisting of 5 components: the CMB anisotropy,
Galactic thermal dust, synchrotron, free-free and anomalous dust
emissions ($j=1,\dots,5$, respectively).

The synchrotron, free-free and anomalous dust components will be
searched at \emph{K}-band frequency $\nu_{0,s}=\nu_{0,ff}=\nu_{0,a}=23\
\rm{GHz}$ and the thermal dust component at \emph{W}-band frequency
$\nu_{0,d}=94\ \rm{GHz}$.  It should be recognized that the Galactic
emission does in general have spectral properties that depend on
position on the sky (especially in the the Galactic plane), however we
will neglect these variations and use a fixed index over the high-latitude 
parts of the sky we are interested in. This is no different
to the assumption that the template fits can be described by single
scale factors over the same region. Even in the case of the \emph{WMAP} ILC method,
the high-latitude emission (as defined by the Kp2 mask) was treated as a
single contiguous region over which the foregrounds were assumed to
exhibit unvarying spectral behaviour. Only within the Galactic plane
as defined by the Kp2 mask were additional regions defined to allow
for spectral variations. 

For the thermal dust component the spectral index generally lies in
the range $1.6
\lesssim \beta_d \lesssim 2.5$ \citep{dupac}. Three values
were selected: $\beta_d=1.7$ suggested by \citet{fds}, $\beta_d=2.2$
which is the mean value of the index determined from the MEM-fitting dust
component derived from the \emph{WMAP} data by \citet{bennet4}, and
the upper limit of the range $\beta_d=2.5$. 

The synchrotron spectral
index is negative and again three values were selected: $\beta_s=-3.3$
which approximately corresponds to the mean spectral index between the
\emph{Ka} and \emph{Q} bands of the \emph{WMAP} MEM-fitting synchrotron component,
$\beta_s=-3.1$ which is a better fit to the Galactic halo emission,
and $\beta_s=-2.9$ more appropriate for the Galactic plane. 

The spectral dependence of the so-called anomalous dust component was
studied by \citet{banday}. Over the frequency range 19 - 90 GHz, the
anomalous dust has a thermal dust-like morphology but a
synchrotron-like spectrum with an index $\beta_a\approx
-2.5$. Unfortunately, the similarity of the latter two spectra results
in the the frequency response matrix becoming poorly conditioned and
the clean separation of the components is not possible. To sidestep
this issue, we adopted the spinning dust model spectrum proposed
by \citet{draine} (hereafter DL98)\footnote{Detailed spectra are available at \hfill\newline
\emph{http://www.astro.princeton.edu/~draine/dust/dust.mwave.spin.html}.}.
In particular, we consider a three-component
model of the spinning dust containing contributions from dust found
in the cold neutral medium (CNM), warm neutral medium (WNM) or warm
ionised medium (WIM). The preferred DL98 model has these 
components in the ratio 0.43, 0.43 and 0.14, respectively.
It should be noted that the model predictions are fundamentally
uncertain because changes in the grain composition can affect 
the turn-over frequency, the spectral normalization, and so forth.
In addition, variations in the mixture of contributions from the
three phases of the interstellar medium can affect the integrated 
spectrum, making our fit a very poor approximation. 
We have tested the sensitivity of the final CMB multipole
reconstruction to the assumptions by replacing the preferred
model mix by either the CNM, WNM or WIM spectra, respectively.
The results are remarkably stable to these variations,
the angle between the quadrupole and octopole axes changing
by less than 3$^\circ$. Furthermore,
\citet{foregroundx} find spectral behaviour in data from 10
-- 90 GHz in reasonable agreement with the predictions of
DL98, and we persist with these assumptions.

Nevertheless, further degeneracy in the components caused by the
similarity of the free-free spectral $\beta_{ff}\approx -2.15$ and the
synchrotron indices, renders it impossible to perform a robust
separation of all 5 components. We have to eliminate the free-free or
synchrotron component from the linear system.  Because the former is
better known (in the sense that there are less variations in the
spectral index over the sky) and of lesser contribution to the
foreground signal, we adopt a free-free template to describe its
spatial morphology and subtract it from the 5 temperature maps,
$\mathbf{d}(\mathbf{r})$, with the correlation coefficients derived by
\citet{bennet4} (as described in \sect\ref{data}).
However, a major uncertainty when using
the H$\alpha$ template is the absorption of the H$\alpha$ by
foreground dust. To study the sensitivity of the lowest multipoles to
the dust absorption correction we will use two templates of the
free-free emission: H$\alpha$ template given by
\citet{finkbeiner} \citep[it was used by ][]{bennet4} uncorrected
for dust absorption and H$\alpha$ template corrected for dust absorption
given by \citet*{dickinson}. Unfortunately, dust may not be the only problem
in removing free-free emission. \citet{finkbeiner2} has suggested that the H$\alpha$ map 
does not trace all of the free-free emission, and proposed the existence of
a so-called free-free haze which has the same spectral dependence as
free-free component. The influence of the free-free haze on the estimation of the lowest
multipoles will be studied using maps corrected and uncorrected for this component.
   
After elimination of the free-free component, an overdetermined linear
system of 5 equations and 4 unknowns remains to be solved.  This is
performed using the singular value decomposition (SVD) method, because
it gives the minimum variance solution. Then, the PE filter was
applied to the derived CMB component maps. Residuals associated with
poorly subtracted Galactic plane emission were circumvented by
application of the Kp2 mask.

The influence of the dust absorption and free-free haze corrections on
the lowest \emph{WMAP} multipoles is shown in
\fig(\ref{cmb_abs_ffhaze}). The dust and synchrotron spectral
indices were fixed here as $\beta_d=2.2$ and $\beta_s=-3.1$,
respectively. One can see that quadrupole is especially sensitive to
these corrections. The dust absorption correction shifts the hotter
spots of the quadrupole toward the Galactic poles and the free-free
haze correction shifts the spots toward the Galactic plane. Curiously,
when both corrections are applied, the quadrupole remains largely
unchanged.  Notable changes are seen also for the $\ell=4$ multipole,
but the results for $\ell=3$ and $5$ are very stable with respect to
the corrections.

The dependence of the $\ell=2$ multipole on the thermal dust and
synchrotron spectral indices is shown in
\fig(\ref{cmb_beta_dust_synch_l2}). We do not show analogous figures
for $\ell=3,4,5$ because they remain unperturbed ($\ell=3,5$) or show
minimal changes ($\ell=4$) with respect to the spectral indices.  It
should be noted that an increase in the synchrotron spectral index
tends to shift the hot spots of the quadrupole toward the Galactic
poles and enhance its amplitude. Conversely, increasing the dust
spectral index shifts the spots toward the Galactic plane and
decreases their amplitude. Thus, for the flattest dust $\beta_d=1.7$
and synchrotron $\beta_s=-2.9$ indices, the quadrupole hot spots are
at the Galactic poles. Similar changes though much smaller are also
seen for multipole $\ell=4$.

The stability of the multipoles $\ell=3$ and $\ell=5$ with respect to
the spectral indices, combined with the earlier results related to the
removal of the free-free template, is compelling evidence in favour of
the claim that the planarity of the octopole is not caused by the
Galactic foregrounds.  Conversely, the strong dependence of the
quadrupole amplitude and morphology on the spectral properties of the
Galactic foreground emission implies that the latter must be known in
some detail in order to recover the quadrupole anisotropy with some
degree of precision.  Indeed, this will also have important
consequences for the statistical significance of its alignment with
octopole. This issue is studied in next section.


\section{The quadrupole and octopole preferred axes} \label{alignment}
As initially noticed by \citet{costa}, and seen here
in \fig(\ref{maps_wmap}), the quadrupole and octopole are rather
planar, with most of their hot and cold spots placed on a single plane in
the sky. de Oliveira-Costa \etal(2004a) 
use a wave function formalism to quantify a preferred axis for the
multipoles and study the significance of their alignment. However, their
analysis was based on the full sky TOH map which may contain residual
contamination in the Galactic plane region. We repeat the
analysis here on the multipoles derived from the cut sky analysis
(thus avoiding this problem), and quantify the additional statistical uncertainties 
introduced by the mask.
 
Given the covariance matrix $\mathbf{M}$, we can estimate uncertainties of the
preferred axes coordinates on the sky. For each of the masks: Kp2, Kp0 and
20+ the covariance matrix $\mathbf{M}$ was used to perform 100 constrained 
realizations of the \emph{WMAP} quadrupole and octopole.
Then, using the same technique as \citet{costa} the preferred axis 
for each of the multipoles was found.  
The uncertainties in the axis coordinates are summarised in \tab(\ref{axes_mask}).      
Notice that the multipoles axes coordinates substantially vary for different
masks. The most changeable is the quadrupole, particularly in latitude, whereas 
the octopole axis coordinates are much more robust.

\begin{table}
\begin{center}
\caption{
	 Galactic coordinates of the quadrupole and octopole axes  
	 for different masks.} 
\label{axes_mask}
{\footnotesize
\begin{tabular}{lcccc}
\hline
 & & Kp2 & Kp0 & 20+ \\
\hline	
$\ell=2$ & Longitude & $273^\circ$ $\pm$ $8^\circ$ & $298^\circ$ $\pm$ $5^\circ$
 & $276^\circ$ $\pm$ $53^\circ$ \\
 & Latitude & $68^\circ$ $\pm$ $3^\circ$ & $40^\circ$ $\pm$ $16^\circ$
 & $34^\circ$ $\pm$ $28^\circ$ \\
$\ell=3$ & Longitude & $307^\circ$ $\pm$ $4^\circ$ & $301^\circ$ $\pm$ $6^\circ$
 & $267^\circ$ $\pm$ $22^\circ$ \\
 & Latitude & $63^\circ$ $\pm$ $2^\circ$ & $62^\circ$ $\pm$ $3^\circ$
 & $66^\circ$ $\pm$ $9^\circ$ \\
 \multicolumn{2}{c}{Angle between axes} & $15^\circ$ $\pm$ $3^\circ$& $22^\circ$ $\pm$ $14^\circ$& $33^\circ$ $\pm$ $23^\circ$ \\
\hline		
\end{tabular}}
\end{center}
\end{table}

\begin{table}
\begin{center}
\caption{
	 Galactic coordinates of the quadrupole and octopole axes for different
	 frequencies. In all cases the multipoles were derived from input maps where 
         the Kp2 mask was applied. The coordinates have uncertainties as in Table 
         (\ref{axes_mask}).} 
\label{axes_freq}
{\footnotesize
\begin{tabular}{lcccccc}
\hline
 & & \emph{K} & \emph{Ka} & \emph{Q} & \emph{V} & \emph{W} \\
\hline	
$\ell=2$ & Longitude & $107^\circ$  & $228^\circ$  & $262^\circ$  & $282^\circ$  & $280^\circ$  \\
& Latitude & $1^\circ$  & $29^\circ$  & $66^\circ$  & $63^\circ$  & $71^\circ$ \\
$\ell=3$ & Longitude & $299^\circ$  & $312^\circ$  & $310^\circ$  & $307^\circ$  & $306^\circ$ \\
& Latitude & $74^\circ$  & $69^\circ$  & $65^\circ$  & $63^\circ$  & $62^\circ$ \\
\multicolumn{2}{c}{Angle between axes}  & $105^\circ$  & $61^\circ$  &  $19^\circ$  &  $11^\circ$   &  $13^\circ$ \\
\hline		
\end{tabular}}
\end{center}
\end{table}

\tab(\ref{axes_freq}) indicates the frequency dependence of the preferred axis for the quadrupole and
octopole.
One should notice that the best foreground corrected
octopole maps (those of the \emph{Q}, \emph{V} and \emph{W} bands) are stable and prefer approximately the same
axis: $(l,b)\sim (308^\circ,63^\circ)$ with a deviation of order $2^\circ$. 
The quadrupole axis is more poorly determined. 
Significant differences between the \emph{Q}, \emph{V}, \emph{W} and \emph{K}, \emph{Ka} frequency bands suggest the
presence of residual foregrounds at the lowest frequencies.   

\begin{table}
\begin{center}
\caption{
	  Galactic coordinates of the quadrupole and octopole axes for
	  four variants of the free-free emission correction of the \emph{WMAP}
	  maps (see text). Columns (1), (2), (3) and (4) correspond to, respectively: no free-free haze
	  correction and no dust absorption correction, free-free haze
	  correction and no dust absorption correction, no free-free haze
	  correction and dust absorption correction, free-free haze
	  correction and dust absorption correction. The dust and synchrotron spectral indices 
	  are $\beta_d=2.2$ and $\beta_s=-3.1$, respectively.} 
\label{tab_dust_hz} 
{\footnotesize
\begin{tabular}{lccccc}
\hline
 & & \multicolumn{2}{c|}{No dust corr.} & \multicolumn{2}{c}{Dust corr.} \\ 	
 & & (1) & (2) & (3) & (4)  \\
\hline		
$\ell=2$ & Longitude&  $297^\circ$  & $292^\circ$  & $300^\circ$  & $296^\circ$  \\ 
 & Latitude&  $53^\circ$  & $69^\circ$  & $33^\circ$  & $62^\circ$ \\ 
$\ell=3$ & Longitude& $306^\circ$  & $301^\circ$  & $301^\circ$  & $299^\circ$\\ 
& Latitude &$62^\circ$  & $62^\circ$  & $62^\circ$  & $60^\circ$ \\
\multicolumn{2}{c}{Angle between axes} & $10^\circ$ &
 $8^\circ$ & $29^\circ$  & $2^\circ$ \\
\hline		
\end{tabular}}
\end{center}
\end{table}

As was shown in \sect\ref{foreground}, the quadrupole of the CMB map derived by 
the linear projection depends significantly on templates used to eliminate the free-free 
emission and assumptions concerning the spectral 
dependence of the dust and synchrotron emission. The coordinates of the quadrupole and 
octopole preferred axes for variants of the free-free model emission are shown in 
\tab(\ref{tab_dust_hz}). Again, the octopole is much more stable than the quadrupole. 
Correction for the putative free-free haze emission
shifts the quadrupole axis toward the Galactic poles
(larger latitude), and the dust 
absorption correction toward the Galactic plane
(smaller latitude).

\begin{table}
\begin{center}
\caption{
	 Galactic coordinates of the quadrupole axes for various
	 spectral indices of the dust and synchrotron emission. The first number in 
	 brackets is longitude and the second latitude. 
	 In all cases the maps were masked by Kp2 mask, thus
	 the coordinates have uncertainties the same as in Table (\ref{axes_mask}) for
	 Kp2 mask.} 
\label{tab_beta}
{\footnotesize
\begin{tabular}{lcccc}
\hline
      & &  \multicolumn{3}{c}{$\beta_{\rm{synch}}$} \\ \hline
      & &  -3.3 & -3.1 & -2.9 \\
\hline	
   & 1.7&  ($298^\circ$, $59^\circ$) & ($298^\circ$, $45^\circ$)
 & ($297^\circ$, $25^\circ$) \\
$\beta_{\rm{dust}}$  & 2.2 & ($291^\circ$, $68^\circ$) & ($296^\circ$, $62^\circ$)
 & ($298^\circ$, $43^\circ$) \\
   & 2.5 & ($287^\circ$, $71^\circ$) & ($294^\circ$, $65^\circ$)
 & ($297^\circ$, $53^\circ$) \\
\hline		
\end{tabular}}
\end{center}
\end{table}

The octopole axis is also stable and does not change for different spectral indices of the dust and synchrotron 
emission. For all studied indices it is $(l,b)=(299^\circ,
60^\circ)$. The quadrupole axis, as in previous cases, is
much less stable (see \tab(\ref{tab_beta})).

In summary of this section it is worth emphasizing that the above results show a much better 
determination of the octopole and its preferred 
axis than for the quadrupole. Significant uncertainties of estimation
for the quadrupole preferred axis and its strong dependence on the details of templates used to 
eliminate free-free emission and foreground spectral indices must increase the low probability
of the \emph{WMAP} quadrupole-octopole alignment estimated by \citet{costa}. The importance of quantifying errors 
as a result of the foreground subtraction
should be clear.


\section{Application to the \emph{COBE}-DMR data} \label{app_cobe}

As a useful cross-check, in particular against systematic effects, we investigate
whether the lowest order multipole maps derived from \emph{COBE}-DMR are consistent
with the \emph{WMAP} results. 
We treated the \emph{COBE}-DMR data in similar way to
the \emph{WMAP} data. Firstly, we corrected the maps for the Galactic foreground using  
the 20+ mask and the Galactic emission templates as was described in
\citet{gorski3} and then co-added the 53 and 90 GHz maps using 
inverse-noise-variance weights. The 31.5 GHz data were not used
because under such a weighting scheme the contribution of these least sensitive
maps is minimal. The PE filter was constructed using the best fit to the 
data power spectrum parameterized by (\citealt{bond}, \citealt*{fabbri}) 
\begin{equation} \label{cl} 
C_\ell\equiv \left<a_{\ell, m}^2\right>=Q_{rms-PS}^2\frac{4
\pi}{5}\frac{\Gamma(\ell+\frac{n-1}{2}) \Gamma(\frac{9-n}{2})}{\Gamma(\ell+
\frac{5-n}{2})\Gamma(\frac{3+n}{2})} \ ,
\end{equation}
where $\Gamma$ denotes the gamma function. The rms quadrupole normalisation 
$Q_{rms-PS}$ and spectral index $n$ were estimated in 
\citet{gorski3}, $Q_{rms-PS}=15.3^{+3.7}_{-2.8}\ \mu \textrm{K}$, $n=1.2\pm 0.3$,
and we used the values of the parameters which maximize the likelihood function
The noise covariance matrix was estimated analogously to the \emph{WMAP} 
matrix. $\ell_{\rm{rec}}$ and $\ell_{\rm{max}}$ were the same as for the \emph{WMAP} 
PE filter. 

The \emph{COBE}-DMR and \emph{WMAP} multipoles maps for the same 20+ mask are compared in 
\fig(\ref{maps_cobe-wmap}). One sees that the maps are similar.  
The \emph{COBE}-DMR octopole appears to be planar  
like the \emph{WMAP} octopole and prefers roughly the same axis, in the direction of
$(l,b)\sim (277^\circ,63^\circ)$. The hot spots of the \emph{COBE}-DMR quadrupole
are shifted more toward the Galactic poles than the spots of the \emph{WMAP} quadrupole,
which may reflect differences in the foreground removal.


\section{Summary} \label{summary}

In this paper, we have re-examined the nature of the low-order multipoles
derived from the \emph{WMAP} data. Previous work has suggested that these
modes demonstrate unusual properties, but such results have been
derived from full sky analyses of maps of the CMB anisotropy derived by methods
which are susceptible to residual Galactic foregrounds \citep{eriksen}, 
particularly in the Galactic plane.
By application of a PE filter on the cut sky we circumvent these difficulties,
and have estimated the amplitudes of the $\ell=2-5$ multipoles and their 
dependence on frequency, sky cut and foreground correction. 
Because the construction of the filter takes into account correlations of the
lower order modes with higher order modes, it significantly improves the accuracy of multipole estimation
in comparison with the direct inversion method used by 
\citet{efstathiou}.

Our studies of the dependence on frequency of the multipole amplitudes 
showed that:
\begin{itemize}
\item[(i)] the Kp2 mask is insufficient to remove strong foreground
contamination near the Galactic plane for the \emph{K} and \emph{Ka} frequency bands even for
the foreground corrected \emph{WMAP} data;
\item[(ii)] small deviations between the $\ell=3,4,5$ multipole determined from the \emph{Q}, \emph{V} and \emph{W}
frequency bands indicates that, after foreground cleaning, these multipoles are 
dominated by the CMB anisotropy component outside the Kp2 mask;
\item[(iii)] the octopole preferred axis, approximately $(l,b)\sim(308^\circ,
63^\circ)$, shows deviations of order two degrees for the \emph{Q}, \emph{V} and \emph{W} frequency
bands;
\item[(iv)] noticeable changes of the quadrupole suggest a non-negligible residual foreground
emission.  
\end{itemize}

That the quadrupole (and to a lesser extent $\ell=4$) may be compromised by such residuals
is supported by the changes observed after the application of increasing
sky cuts. It also appears that the estimation of the quadrupole is
sensitive to the dust absorption  
correction of the free-free template and free-free haze correction. This
may provide yet further independent 
confirmation of significant contamination of the quadrupole by residual
Galactic foregrounds.  

Further tests of the sensitivity of the multipoles to the spectral dependence of the foreground emission  
have utilised a linear projection technique for foreground removal. 
Studies of the dependence of the multipole amplitudes and morphologies on
the dust and synchrotron spectral indices showed that:  
\begin{itemize} 
\item[(i)] multipoles $\ell=3,4,5$ are independent of the indices and do not
vary with respect to them;   
\item[(ii)] increase of the synchrotron spectral index shifts hotter spots toward 
the Galactic poles and enhances amplitude of the quadrupole;   
\item[(iii)] increase of the dust spectral index shifts hotter spots toward the  
Galactic plane and decreases amplitude of the quadrupole. 
\end{itemize} 

In conclusion, it remains premature to make definitive statements in
relation to the quadrupole. Large uncertainties of the quadrupole preferred axis should
substantially increase the very low 
probabilities of the \emph{WMAP} quadrupole-octopole alignment estimated by
\citet{costa} and other quadrupole correlations studied by
\citet{schwarz}. To similar conclusion came also \citet{slosar} who showed
that much of the alignment between quadrupole and octopole is the result of foreground contamination.   
Nevertheless, estimates of $\ell=3,4,5$ are quite robust, and consistent
with previous values. The observed anisotropy structures are also present for in \emph{COBE}-DMR data,
and are therefore unlikely to be associated with instrumental systematic artifacts. 
  
\section{Acknowledgments} 

We acknowledge the use of CMBFAST \citep{cmbfast}. 
Some of the results in this paper have been derived using the HEALPix
\citep{healpix} software and analysis package.
We acknowledge the use of the Legacy Archive for Microwave Background
Data Analysis (LAMBDA). Support for LAMBDA is provided by the NASA
Office of Space Science.
PB acknowledges the financial support provided through the European
Community's Human Potential Programme under contract HPRN-CT-2000-00124,
(CMBNET), and Max-Planck-Institut fuer Astrophysik in Garching where parts
of this work were carried out.



\begin{appendix}

 
\section{Covariance matrix} \label{covmat}

The posterior probability is a multivariate Gaussian of the form
\begin{equation} \label{post_prob}
\textrm{Pr}(\mathbf{a}_{\mathcal{L}}|\mathbf{c})\propto \exp
\left(-(\mathbf{a}_{\mathcal{L}}-\hat{\mathbf{a}}_{\mathcal{L}})^T \cdot
\mathbf{M}^{-1} \cdot (\mathbf{a}_{\mathcal{L}}-\hat{\mathbf{a}}_{\mathcal{L}})/2
\right)/\sqrt{\det{\mathbf{M}}}\ , 
\end{equation}
which has its maximum value at the estimate
$\hat{\mathbf{a}}_{\mathcal{L}}$ of the signal vector and
$\mathbf{M}$ is the covariance matrix defined
by $\mathbf{M} \equiv \left< \left( \mathbf{a}_\mathcal{L}-\hat{\mathbf{a}}_\mathcal{L} \right) \left(
\mathbf{a}_\mathcal{L}-\hat{\mathbf{a}}_\mathcal{L} \right)^T
\right>$. The diagonal terms of the covariance matrix $\mathbf{M}$ are the squared
errors $\Delta \mathbf{a}_\mathcal{L}$ of the estimate
$\hat{\mathbf{a}}_\mathcal{L}$ and non-diagonal elements of the matrix represent correlations
between errors. However, to study correlations a more convenient representation of the
correlation matrix is used -- the covariance matrix normalized by its diagonal terms, $\mathbf{M}_{ij}/
\sqrt{\mathbf{M}_{ii} \mathbf{M}_{jj}}$
( $i,j\in[5,\dots,(\ell_{\rm{rec}}+1)^2]$ ). Then, the coefficient of
correlation between
uncertainty in $\mathbf{a}_i$ and uncertainty in $\mathbf{a}_j$ is a number between -1 and 1.
\end{appendix}

\begin{figure*}
\vspace{102pt}
\caption{Maps of the \emph{WMAP} $\ell=2,\dots,5$ multipoles, computed
by the PE filtering of the \emph{WMAP} corrected
co-added map (first row), TOH cleaned map (second row) and ILC map
(third row) after applying the Kp2 mask. 
} 
\label{maps_wmap}
\end{figure*}

\begin{figure*}
\vspace{102pt}
\caption{Maps of the $\ell=2,\dots,5$  \emph{WMAP} multipoles determined from the 
different frequency bands.} \label{maps_wmap_freq}
\end{figure*}

\begin{figure*}  
\vspace{102pt}
\caption{Maps of the $\ell=2,\dots,5$  \emph{WMAP} multipoles for
different masks: Kp2, Kp0, 20+ and 30.} \label{maps_wmap_cut}
\end{figure*}

\begin{figure*} 
\vspace{80pt}
\caption{The $\Delta_{\rm{error}}^2$ maps (see text) corresponding to the sum of 
the errors for the restored $\ell =2,\dots,5$ \emph{WMAP} multipoles (centre)
and for each multipole separately. The maps are shown in logarithmic scale.}
\label{map_st_dev_wmap}
\end{figure*}
 
\begin{figure*}
\vspace{50pt}
\caption{The multipoles map obtained by the PE filtering of the CMB maps after masking by
the Kp2 mask. The CMB maps were derived from the \emph{WMAP} data as corrected by four 
variants for the free-free emission. From top to bottom, the maps 
for following variants of the correction are shown: for H$\alpha$ template uncorrected 
for dust absorption and no free-free haze correction, for H$\alpha$ template  
corrected for dust absorption and no free-free haze correction, for H$\alpha$ template 
uncorrected for dust absorption and free-free haze correction, for H$\alpha$ template 
corrected for dust absorption and free-free haze correction. Spectral indices of the thermal 
dust and synchrotron emission were $\beta_d=2.2$ and $\beta_s=-3.1$, respectively.}
\label{cmb_abs_ffhaze}
\end{figure*} 

\begin{figure*}
\vspace{50pt}
\caption{The $\ell=2$ quadrupole component obtained by PE filtering
of the CMB maps after masking by the Kp2 mask. 
The CMB maps were derived from the \emph{WMAP} maps corrected for the free-free 
emission (with H$\alpha$ template corrected for dust absorption and free-free haze template) 
for various spectral indices of the dust $\beta_{dust}$ and synchrotron 
$\beta_{sych}$ indices.}
\label{cmb_beta_dust_synch_l2}
\end{figure*} 

\begin{figure*}
\vspace{50pt}
\caption{Maps of the $\ell=2,\dots,5$ multipoles recovered using the
PE filter for \emph{COBE}-DMR (the first column) and \emph{WMAP} data (the second column).  
In both cases the extended Galactic cut was used.}
\label{maps_cobe-wmap} 
\end{figure*}

\end{document}